\documentclass{aa}
\usepackage[labelfont=bf,textfont=footnotesize]{caption}  

\usepackage[varg]{txfonts}
\usepackage{epsfig,graphicx,natbib,url,twoopt}
\usepackage[varg]{txfonts}
\usepackage[colorlinks=true,urlcolor=blue,citecolor=blue,linkcolor=blue,breaklinks=true]{hyperref}
\usepackage{gensymb}

\let\oldhref\href
\renewcommand{\href}[2]{\oldhref{#1}{\hbox{#2}}}

\bibpunct{(}{)}{;}{a}{}{,} 
\makeatletter
\newcommandtwoopt{\citeads}[3][][]{\href{http://adsabs.harvard.edu/abs/#3}%
{\def\hyper@linkstart##1##2{}%
\let\hyper@linkend\@empty\citealp[#1][#2]{#3}}}
\newcommandtwoopt{\citepads}[3][][]{\href{http://adsabs.harvard.edu/abs/#3}%
{\def\hyper@linkstart##1##2{}%
\let\hyper@linkend\@empty\citep[#1][#2]{#3}}}
\newcommandtwoopt{\citetads}[3][][]{\href{http://adsabs.harvard.edu/abs/#3}%
{\def\hyper@linkstart##1##2{}%
\let\hyper@linkend\@empty\citet[#1][#2]{#3}}}
\newcommandtwoopt{\citeyearads}[3][][]%
{\href{http://adsabs.harvard.edu/abs/#3}
{\def\hyper@linkstart##1##2{}%
\let\hyper@linkend\@empty\citeyear[#1][#2]{#3}}}
\makeatother

\newcommand{\bq}{\begin{equation}}
\newcommand{\eq}{\end{equation}}
\newcommand{\beaq}{\begin{eqnarray*}}
\newcommand{\eeaq}{\end{eqnarray*}}

\newcommand{\efft}{\ion{Ca}{ii}~8542~\AA\ }
\newcommand{\cahline}{\ion{Ca}{ii}~H line}
\usepackage{epstopdf}
\usepackage{epsfig}
\usepackage{threeparttable}

\begin{document}  

\title{Microjets in the penumbra of a sunspot}

\author{Ainar Drews
\and
Luc Rouppe van der Voort}

\institute{Institute of Theoretical Astrophysics, University of Oslo, P.O. Box 1029
Blindern, N-0315 Oslo, Norway, \email{ainar.drews@astro.uio.no}}

\date{\today}

\abstract
 {
 Penumbral Microjets (PMJs) are short-lived jets found in the penumbra of sunspots, first observed in wide-band \cahline\ observations as localized brightenings, and are 
 thought to be caused by magnetic reconnection. 
 }
 {
 Earlier work on PMJs has been focused on smaller samples of by-eye selected events and case studies. It is our goal to present an automated study of a 
 large sample of PMJs to place the basic statistics of PMJs on a sure footing and 
 to study the PMJ \efft spectral profile in detail.  
 }
 {
 High spatial resolution and spectrally well-sampled observations 
 in the \efft line obtained from the Swedish 1-m Solar Telescope (SST) are reduced by a Principle 
 Component Analysis and subsequently used in the automated detection of PMJs using the simple learning algorithm k-Nearest Neighbour.  
 PMJ detections were verified with co-temporal \cahline\ observations. 
 }
 {
 A total of 453 tracked PMJ events were found, or 4253 PMJs detections tallied over all timeframes and a detection rate of 21 events per timestep. From these, an average length, width and lifetime of 640~km, 210~km and 90~s were obtained. 
 The average PMJ \efft line profile is characterized by enhanced inner wings, often in the form of one or two distinct peaks, and a brighter line core as compared to the quiet Sun average. 
Average blue and red peak positions were determined at $-10.4$~km~s$^{-1}$ and $+10.2$~km~s$^{-1}$ offsets from the \efft line core.  
 We found several clusters of PMJ hotspots within the sunspot penumbra, where PMJ events occur in the same general area repeatedly over time.
 }
 {
 Our results indicate smaller average PMJs sizes and longer lifetimes compared to previously published values, but with statistics still in the same orders of magnitude. 
 The investigation and analysis of the PMJ line profiles strengthen the proposed heating of PMJs to transition region temperatures. The presented statistics on PMJs form a solid basis for 
 future investigations and numerical modeling of PMJs.  
 }

\maketitle

\section{Introduction \label{sec:introduction}}     

Penumbral microjets (PMJs) are short-lived, elongated, transients in the chromosphere of sunspot penumbrae.
They were discovered 
\citepads{2007Sci...318.1594K}  
in Ca II H time sequences from Hinode's 3~\AA\ wide imaging filter
in which PMJs display a 10--20\% brightness enhancement as compared to surrounding penumbral structures.
In the Hinode observations, PMJs have typical lifetimes of up to 1 minute, lengths between 1000 and 4000~km, widths of about 400~km, and apparent rise velocity faster than 100~km~$^{-1}$ 
\citepads{2007Sci...318.1594K}. 
\newline
Penumbrae are known to host strong convectively driven plasma flows and magnetic fields that vary significantly at small spatial scales, both in inclination and magnitude (see e.g., 
\citeads{2011LRSP....8....4B}).  
In this magnetically stressed environment, magnetic reconnection appears a viable candidate as driver of PMJs.
This is supported by the measurements of the apparent inclination of PMJs with respect to the photospheric penumbral filaments 
\citepads{2007Sci...318.1594K}, 
and magnetic fields 
\citepads{2008A&A...488L..33J}.  
Further indications of the reconnection scenario come from 
\citeads{2010A&A...524A..20K}  
who reported the association of small photospheric downflow patches with some PMJs.
These could be interpreted as the downward flows from magnetic reconnection above the photosphere.
Evidence of progressive heating along PMJs was reported by
\citeads{2015ApJ...811L..33V} (Fig.5), 
who found clear responses in \ion{Mg}{ii} k, \ion{C}{ii}, and \ion{Si}{IV} slit-jaw images of IRIS to PMJs observed in Ca II lines.
\ion{C}{ii} and \ion{Si}{IV} emission towards the top of PMJs suggests heating to transition region temperatures.
\newline
\citeads{2013ApJ...779..143R}   
study transients in a sunspot penumbra from spectral imaging data in the Ca II 8542 line.
Aided by co-temporal Hinode Ca II H imaging, they identified several PMJs in their dataset. The Ca II 8542 line profiles show enhanced emission in the wings out to $\pm0.5$~\AA, with peaks at about $\pm0.3$~\AA, and a line core that shows little difference as compared to the surroundings.
They point at the similarity with Ca II 8542 spectral profiles of Ellerman bombs, like for example as shown in 
\citeads{2013ApJ...774...32V}. 
\newline
In this study, we expand on the observational characterization of PMJs analyzing a high-spatial resolution time series of both narrow-band (1~\AA) Ca II H filtergrams and spectral imagery in Ca II 8542.
We employ an automated detection scheme to built a large statistical sample of PMJs.
The detection scheme neatly takes advantage of the many sampling position in the Ca II 8542 line, utilizing the full line profile. The dimensionality of the observations in wavelengths positions is first reduced employing Principal Component Analysis (PCA),  
and is then used in detections performed employing the k-Nearest Neighbour algorithm,  
and finally followed by object tracking and statistical analysis.

\section{Observations \label{sec:obs}}

\begin{figure*}[!htb]
\centering
\includegraphics[height=7.0cm]{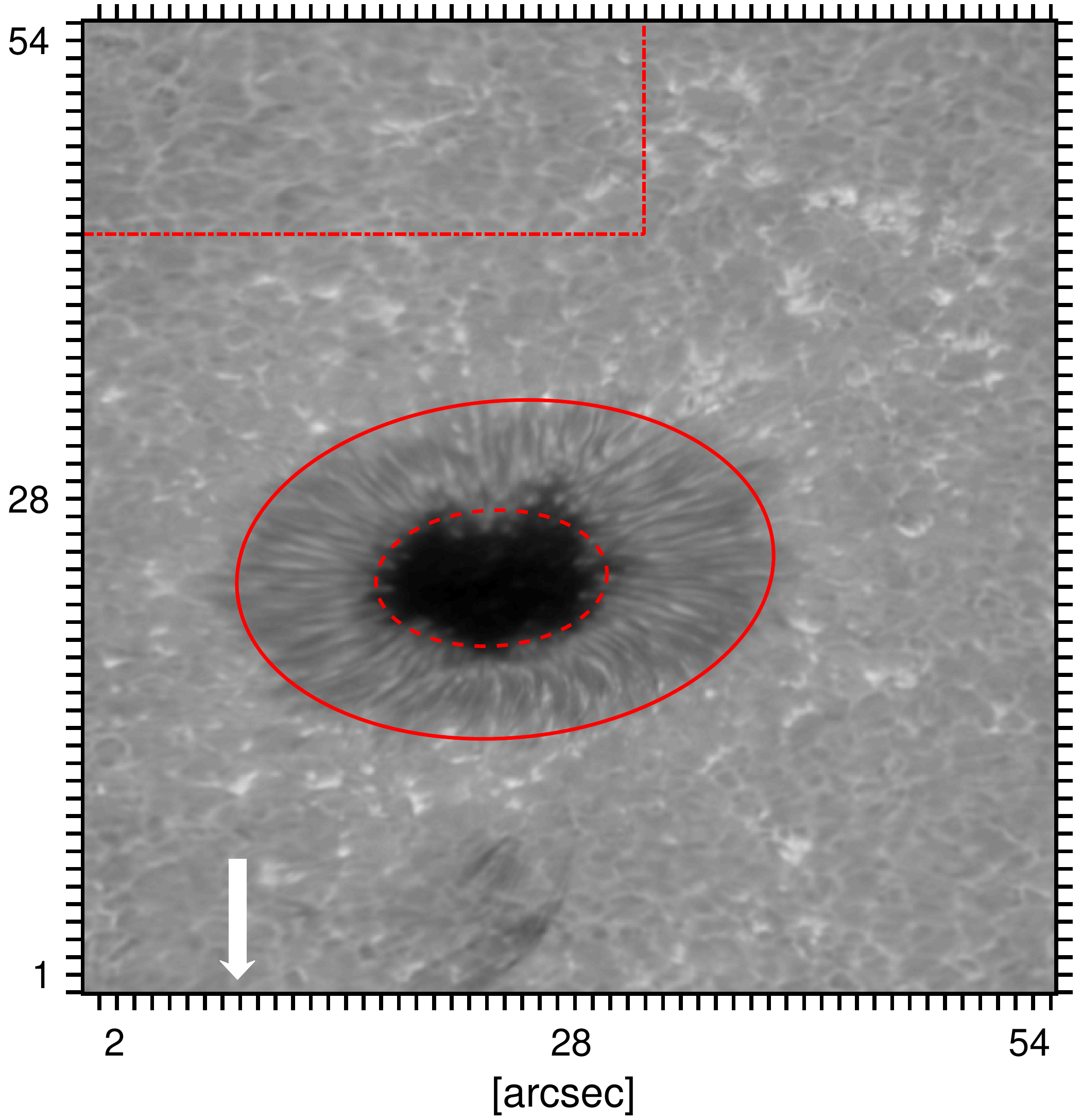} 
\includegraphics[height=5.6cm]{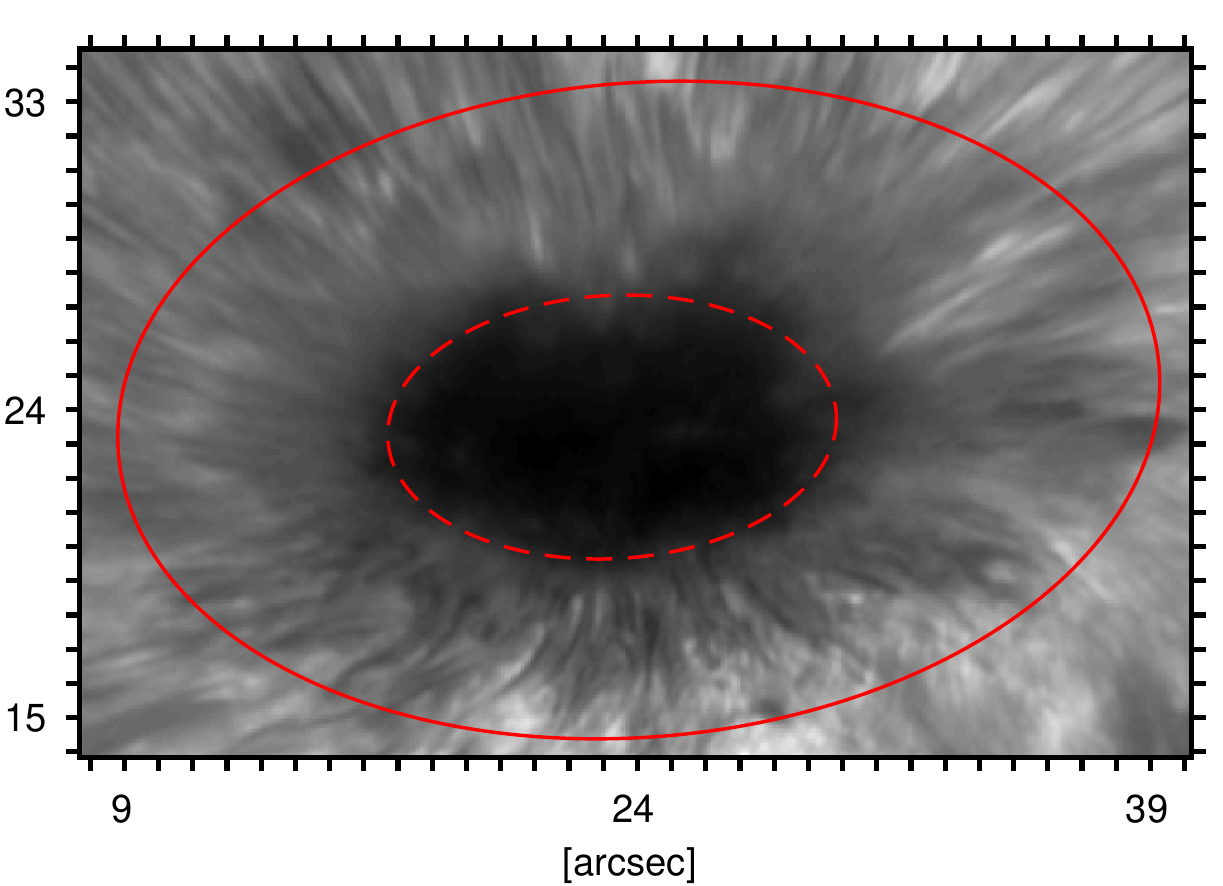} 
\includegraphics[height=6.4cm]{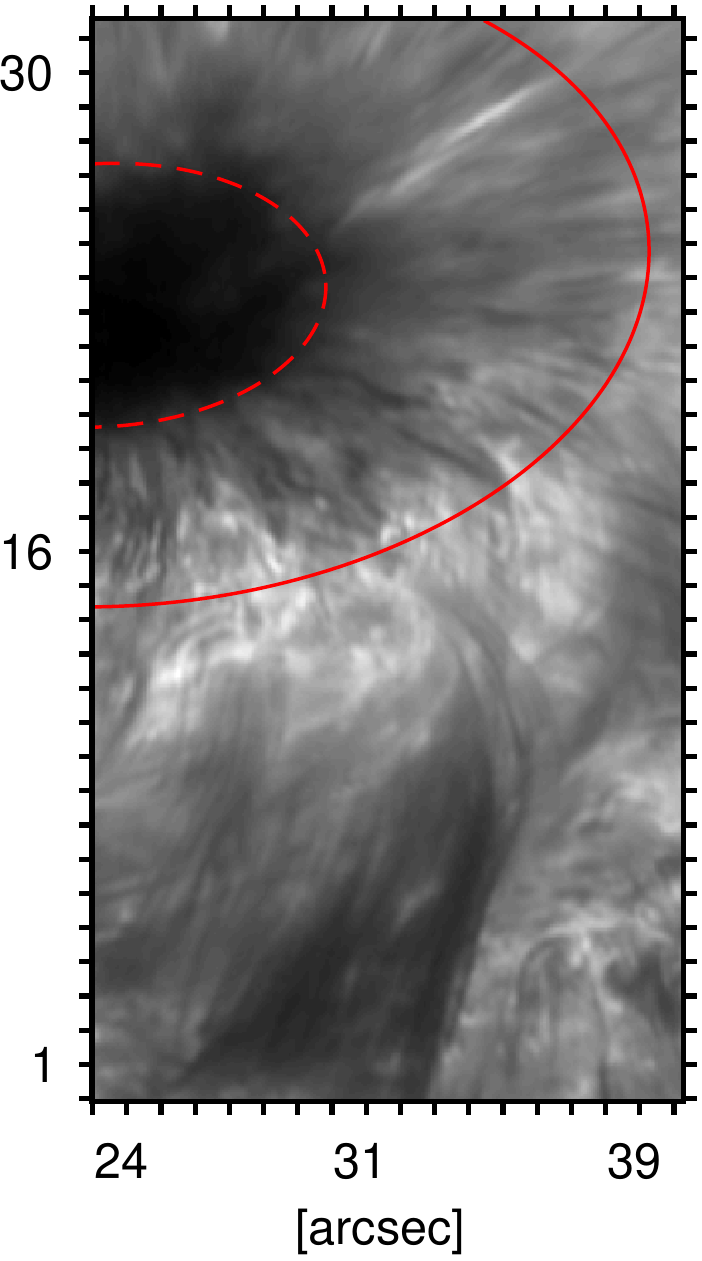} 
\caption{\label{fig:sunspot_ffov}Sunspot imaged by the SST on 28-June-2010 in the \efft\  line at offsets $-1032$~m\AA\ in the full field of view (left) and $-275$~m\AA\ (middle and right) in the \efft\ line, zoomed in on the  
(middle) and a subfield that contains a strong inverse Evershed 
flow (the dark feature extending from the bottom of the frame to the edge of the penumbra) (right). The borders of the umbra (dashed-red) and penumbra (solid-red) are indicated. Also indicated (left) is the area over which the
reference quiet sun \efft\ average line-profile was computed (dash-dotted-red). The arrow (solid-white) indicates the direction towards disk-center. Images (left) and (middle) are at time 23 min 27 s, image (right) is at 37 min 46 s.  
}
\end{figure*}

Active region AR11084 was observed on 28-June-2010 with the Swedish 1-m Solar 
Telescope 
\citepads[SST, ][]{2003SPIE.4853..341S} 
on La Palma. The field of view was centered on the near-circular sunspot with 
fully developed penumbra at heliocentric coordinates $(X,Y)=(710,-339)$ 
($\mu = \cos \theta = 0.55$, with $\theta$ the observing angle). The seeing was
 excellent for the full 41 min duration of the time series
 (starting at 09:18:29 UT), and the image quality further benefited from the 
adaptive optics system 
\citepads{2003SPIE.4853..370S} 
and image reconstruction with the Multi-Object Multi-Frame Blind Deconvolution
 method
\citepads[MOMFBD, ][]{2005SoPh..228..191V}. 
We analyzed data from instruments on both branches of the optical beam: from the
 CRisp Imaging SpectroPolarimeter
\citepads[CRISP, ][]{2008ApJ...689L..69S} 
on the long-wavelength branch ``red beam''), and filtergram imaging in 
\ion{Ca}{ii}~H on the short-wavelength branch (``blue beam'').

With CRISP, we sampled the \ion{Ca}{ii}~8542~\AA\ line at 37 line positions, 
with equidistant 55~m\AA\ steps out to $\pm880$~m\AA, and additional sampling 
at $\pm948$ and $\pm1034$~m\AA. CRISP has  a FWHM of 110~m\AA\ at 8542~\AA, so 
the \ion{Ca}{ii} line is critically sampled throughout the central part of the 
spectral line profile. We acquired 8 exposures per spectral sampling which were 
used for MOMFBD image reconstruction. In addition, single wavelength 
spectro-polarimetric samplings of the \ion{Fe}{i}~6302~\AA\ line were acquired. 
Unfortunately, due to erroneous calibration settings, the precise wavelength for 
this sampling was unknown which resulted in noisy and effectively useless maps
 of the four Stokes parameters. The acquisition time for \ion{Ca}{ii}~8542~\AA\
 was 8.1~s and the temporal cadence of the time series was 12.4~s. After MOMFBD 
restoration of the individual spectral line scans, the data was put together as
 a time series after correction for the CRISP prefilter (FWHM 9.3~\AA\ for 
\ion{Ca}{ii}), compensation of the diurnal field rotation, rigid alignment, and 
destretching.We used early versions of the different procedures that were later 
put together as the reduction pipeline for CRISP data 
\citepads{2015A&A...573A..40D} 
including the post-MOMFBD correction for remaining small-scale seeing 
deformations due to the non-simultaneity of the sequentially recorded narrowband
 CRISP images 
\citepads{2012A&A...548A.114H}. 
The effective field of view of the time series is 55\arcsec $\times$ 55\arcsec, 
with a pixelscale of 0\farcs059 per pixel. 

In the blue beam, synchronized filtergrams were recorded at a rate of 10.8 
frames s$^{-1}$ in the \ion{Ca}{ii}~H line core (filter FWHM 1.1~\AA) and with
 a wider passband filter (FWHM 10~\AA) at $\lambda=3954$~\AA, between the 
\ion{Ca}{ii}~H and K lines. These two imaging channels were MOMFBD restored to 
produce a time series with an effective cadence of half the CRISP data, 6.2~s. 
The alignment to the CRISP data was done by cross-correlation of the red and
 blue wideband channels, which both show the photosphere. 

Figure~\ref{fig:sunspot_ffov} shows example frames from the observations, 
displaying the sunspot at offsets $-1032$~m\AA\  and $-275$~m\AA\ in the \efft\ line. Also 
displayed are the nominal borders of the umbra and penumbra as well as the area over which 
the average quiet sun \efft\ line-profile used for comparison was computed over. A cropped 
image of the observations also showcases a strong inverse Evershed flow.


\section{Methods \label{sec:methods}}   

Here a brief overview of the used methods is given. Prior to the employment of the 
pipeline described below, the observations were investigated using the CRisp SPectral EXplorer
\citepads[CRISPEX]{2012ApJ...750...22V}, 
which was used to interactively browse the MOMFBD reduced observations. CRISPEX was also 
instrumental in the by-eye assembly of the reference set of PMJs and non-PMJ objects in the 
observations for later employment of the k-NN algorithm.

We restrict to a qualitative description and focus only on the basic concepts and overall structure. 
For an in-depth treatment, the reader is directed to 
\cite{Drews}, 
Section 5, in which the full 
methodology for the automated detection scheme and detection process is described, as well 
as a full explanation of the subsequent object-tracking. 

The main working steps, starting from the post-MOMFBD
SST observations, can be summarized in the following discrete steps: 

\begin{enumerate} 
\item Preliminary identification of PMJs in \efft\ using \cahline\ observations as reference
\item Principle Component Analysis: dimensionality reduction and data compression 
\item Detection of PMJs using the k-Nearest Neighbour algorithm
\item Object tracking and statistical analysis
\end{enumerate} 

These four distinct steps are covered in some more detail in the subsections below. 

\subsection{Preliminary identifications of PMJs}

To justify the claimed observation of PMJs in the \efft\ line, 
a subset of 
observations was first compared to by-eye detections of PMJs in the 
co-observed \cahline, as the detection of PMJs and their appearance is firmly 
established in this line 
(\citeads{2007Sci...318.1594K}, 
\citeads{2008A&A...488L..33J},  
\citeads{2010A&A...524A..20K}  
and 
\citeads{2010A&A...524A..21J}). 

Earlier observations of PMJs in \ion{Ca}{ii}~8542~\AA\ were presented by
\citeads{2013ApJ...779..143R},  
\cite{Drews},  
as well as by
\citeads{2015ApJ...811L..33V}.

A quick qualitative study was enough to show that many, if not most, \ion{Ca}{ii}~H PMJ
detections have spatially coinciding similar features in the \efft\ line,
in particular slightly blue-ward of the nominal line center wavelength. 
Similar as in \ion{Ca}{ii}~H, PMJs appear in 
selected spectral positions in the \ion{Ca}{ii}~8542~\AA\ line as short-lived, 
elongated brightenings in the sunspot penumbra. 

\begin{figure*}[!tb]
\includegraphics[width=\columnwidth]{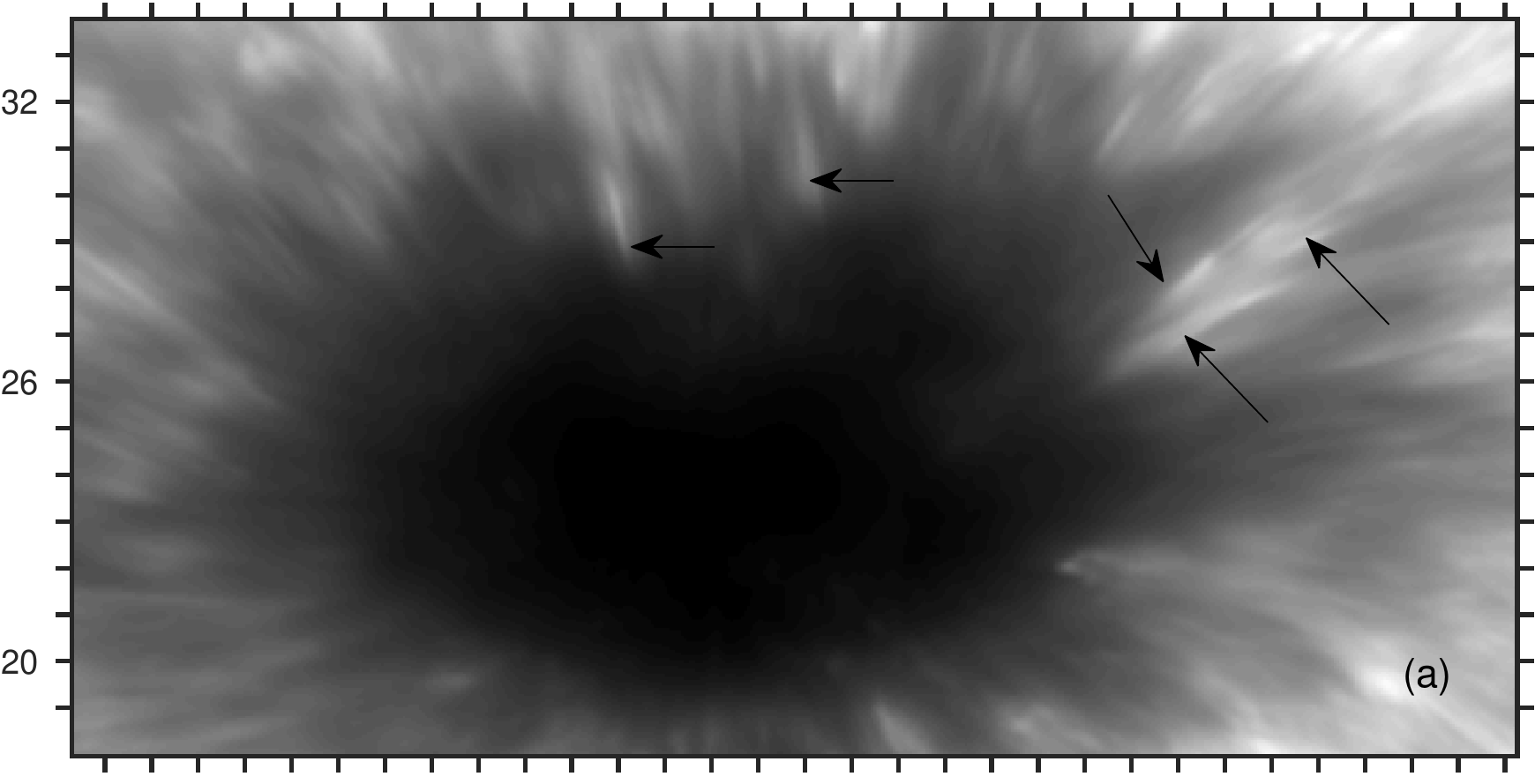} 
\includegraphics[width=\columnwidth]{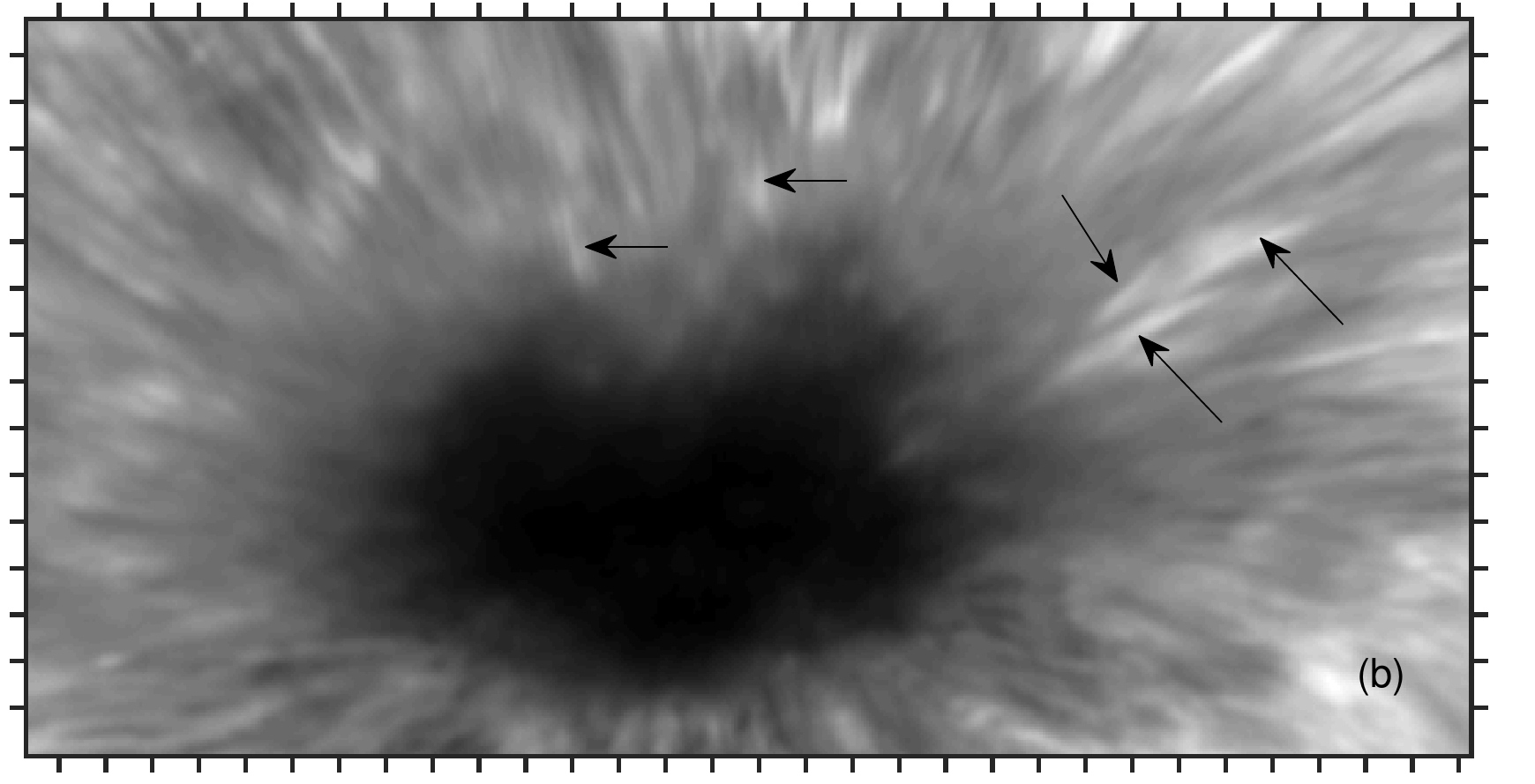} \\
\includegraphics[width=\columnwidth]{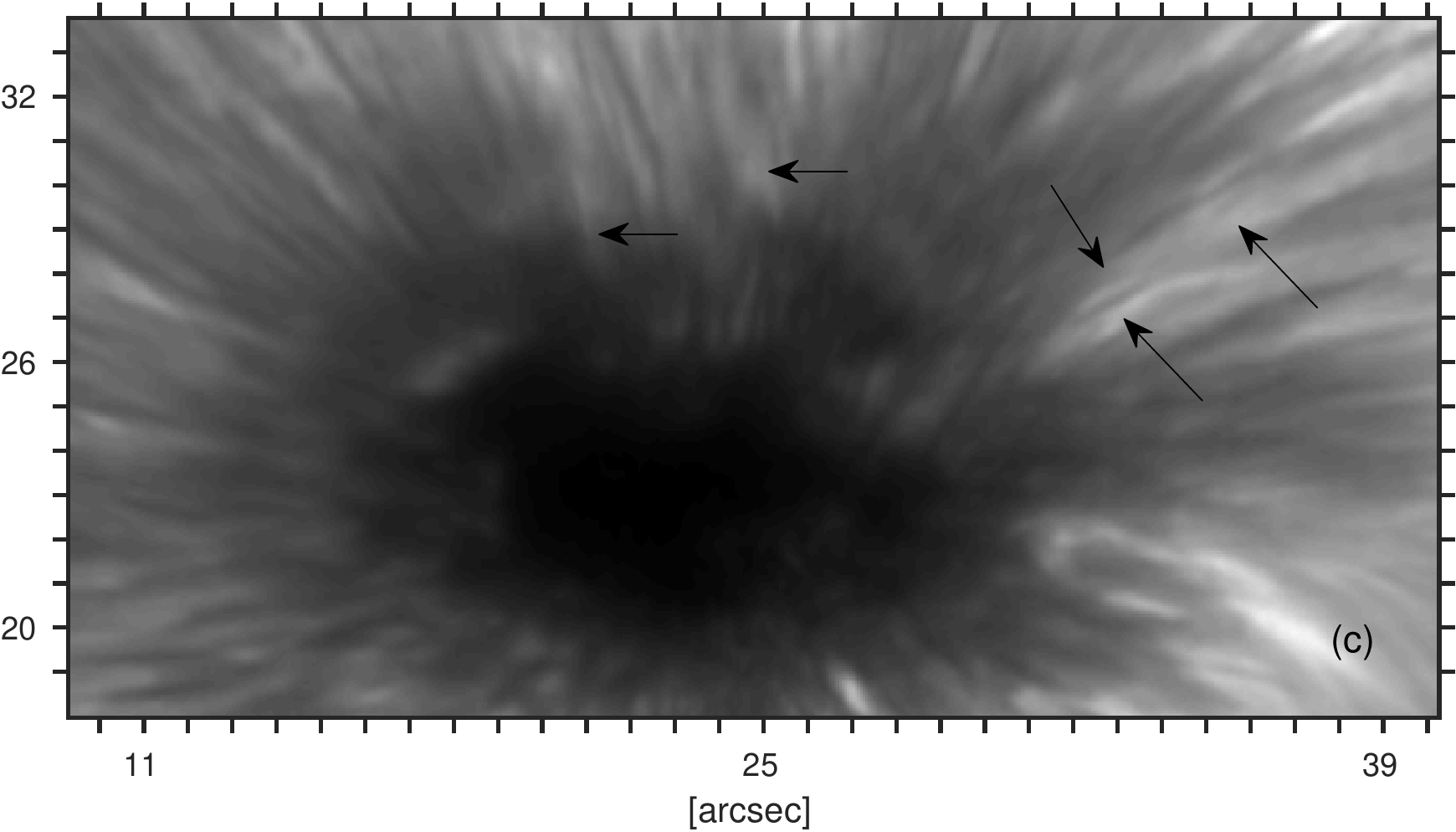} 
\includegraphics[width=\columnwidth]{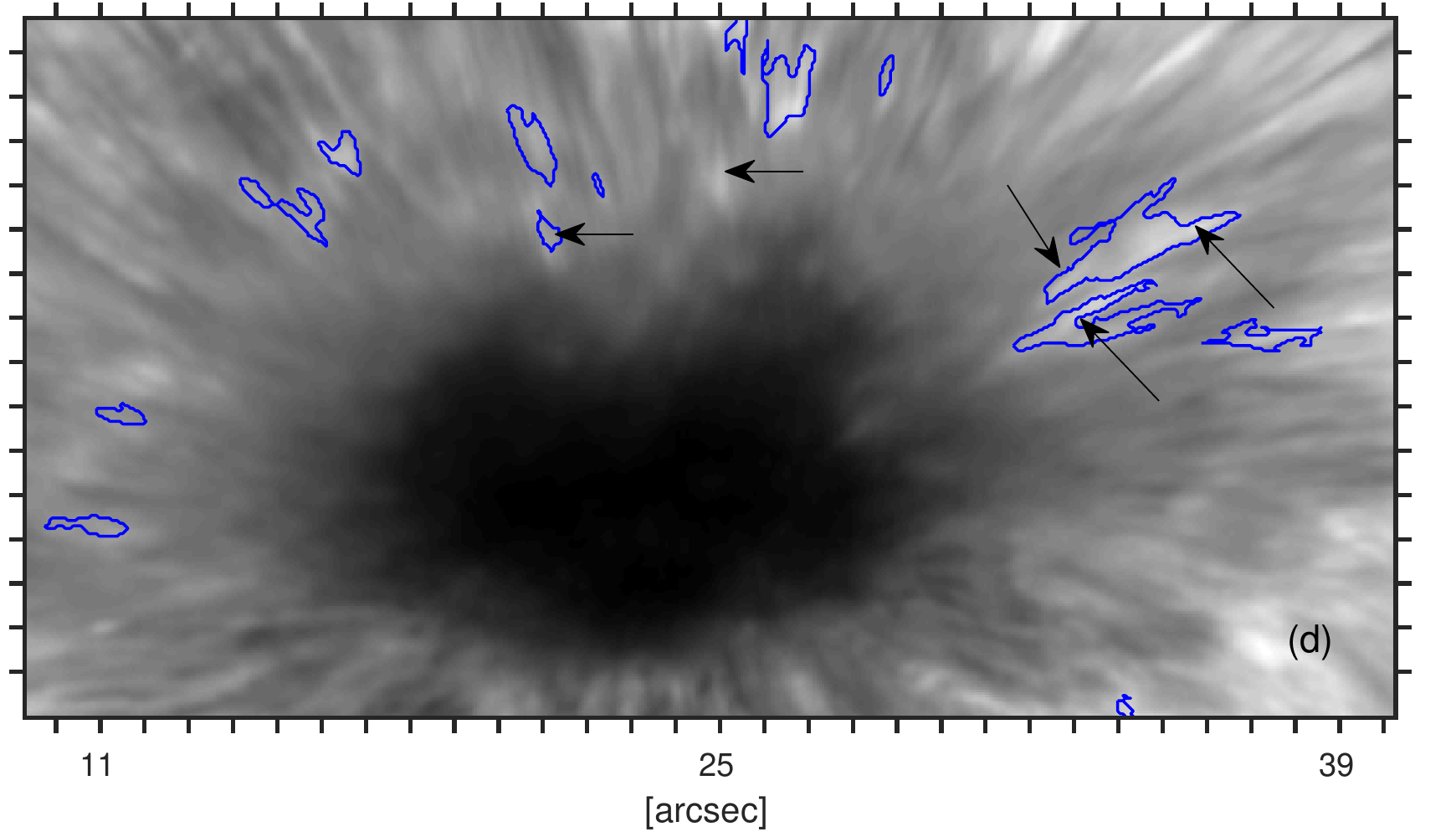} 
\caption{\label{fig:prev_H_22}
Different wavelengths at time 25 min 6 s, with arrows at identical pixel positions indicating PMJ events 
(a) \ion{Ca}{ii}~H line core (FWHM 1.1 $\AA$) 
(b) \ion{Ca}{ii}~8542 at $-275 $~m\AA\ offset from line core (FWHM 110 m$\AA $ in the \ion{Ca}{ii}~8542 line) 
(c) \ion{Ca}{ii}~8542 line core 
(d) \ion{Ca}{ii}~8542 at $-275 $~m\AA\ offset from line core with the detection borders of automatically detected PMJs in the FOV overplotted (blue-solid).
}
\end{figure*}

This is illustrated in Fig.~\ref{fig:prev_H_22}, which shows four example images in the observations at the 
same timeframe. One image is in the \cahline, one in the \ion{Ca}{ii}~8542 line core and two in the \efft line 
at an offset of $-275$~m\AA. PMJs proved to be most visible in by-eye detections in the \efft line in images at an offset of $-275$~m\AA. 
The arrows in the four panels all point to the same pixel positions, and it is clear that PMJ 
features are present in both diagnostics (the \cahline\ and the \efft line scan). This is especially true 
when comparing features present in the \cahline\, which are clearly visible in the $-275$~m\AA\ \efft\ line offset images as well. 
Detection borders of PMJs overplotted in panel (d) highlight that most by-eye selected examples in this 
frame were caught by the automated detection scheme that will be presented below (see Sect. \ref{sec:PCA} - \ref{sec:computing_profile}). 
Notably, two PMJs as selected by-eye on the right of the FOV are classified as one event by the automated scheme. 
Further, one of the events is not detected by the detection scheme, marked by the second arrow from the left (manual inspection using 
CRISPEX indicate that this event's spectral profile is not very distinct over a larger area).  

From these examples, and other qualitative inspections of PMJs that visually coincided spatially in the \cahline\ as 
well as in the \efft\ line observations, the assumption that these events are the same physical objects 
is validated. Further investigation of PMJs in the \efft\ line and the ascertainment of a distinct spectral line profile 
is therefore warranted.

\subsection{Principle Component Analysis}\label{sec:PCA}

The detections of PMJs in the presented automated approach is based on their distinct 
line profiles in the \efft\ line. Intensity differences throughout the field of view, due to 
both limb-darkening as well as local variations, proved to make detections using the k-NN algorithm 
difficult. Intensity variations affecting the whole line profile may not impact the shape of 
the given pixel's line profile, but may shift the overall intensity, including key sampling points where PMJs are 
visible, such that the similarity measures in the algorithm were less meaningful. For this reason, prior to PCA treatment, the observations 
were first normalized. The normalization consisted of normalizing each line profile in all pixels in all timeframes 
to its own sum. This preserved the line profile shapes, but effectively removed any overall intensity 
variations. The preservation of line profiles shapes consequently carried over into the PCA treated observations used 
in the k-NN detection scheme. 

PCA is both a method for data analysis, as well as 
being a useful tool for data compression. 
In \cite{DBLP:journals/corr/Shlens14} 
an introductory overview to PCA is given. 

In PCA, the covariance between variables in a dataset is computed, constructing 
its covariance matrix. Subsequently, the associated eigen-vectors and -values 
of this matrix are found. The eigenvectors will correspond to a new set of basis 
vectors along which the data can be projected, yielding a linearly independent 
dataset with no cross-correlation between variables. Normalizing the associated 
eigenvalues, these yield the relative contribution of the new basis vector 
variables to the total variance of the newly aligned dataset. In practice, this 
variance can be equated to the informational contribution from the given new 
eigenvector. This presents the opportunity for compression of the dataset, as 
discarding variables along eigenvectors with low informational value will not 
yield significant information loss overall, and additionally, this loss will be 
quantifiable. 

In the present context, the 37 wavelength sample points of the \efft\ line are treated as 
the variables in a 37-dimensional dataset. Hence, this is the dataset that is 
analyzed and compressed using PCA as outlined above. 

For the present pipeline, the computation of the covariance matrix 
of the \efft\ line observations was performed following 
\cite{Bennet_et_al}. 

Here, a numerically stable single-pass algorithm is presented. It provides the benefit 
of a lighter numerical workload as a naive (but generally stable) approach requires two passes over 
the dataset. In such an approach the mean of a dataset is first computed, followed by the computation of the needed 
powers of this mean in the second pass. 

The employed single-pass of \cite{Bennet_et_al}
also avoids common numerical pitfalls in the computation of the covariances in terms of numerical instabilities when 
calculated in a single-pass approach. 
 
Using the found eigenvectors and corresponding eigenvalues, it was determined 
that the variables along 7 eigenvectors describe the original observations 
to an accuracy, or informational content, of 97\%, which was deemed acceptable, 
whilst yielding a compression of the data to 19\% of the original size. Thus, 
these 7 eigenvectors were chosen as the Principle Components of the dataset, 
and the observations aligned and compressed along them.

Different morphological features, all with distinct spectral profiles (e.g., long fibrils, umbral flashes, fibrils with strong flows) are clearly identifiable in maps of the different Principle Components.
This lends intuitive credence to the PCA reduction method, in that the different
 ``new spectral'' variables still represent real features picked out from the 
original sampling positions.
For examples, we refer to \cite{Drews},  
Section 5.

\subsection{The k-Nearest Neighbour algorithm}

The k-Nearest Neighbour 
algorithm is conceptually simple, yet powerful. This is one of 
the reasons it is widely applied in signal processing tasks such as facial and voice 
recognition as well as machine reading. For an introduction to the specifics of the 
algorithm, an overview and discussion for improvement of the algorithm is given in 
\cite{Guo_et_al}. 
Further, 
\cite{Yang_et_al} 
compare the algorithm to other classifiers. 

Often termed the simplest of the learning algorithms, the k-NN algorithm is based on a 
comparative approach in which data to be classified is related to a pre-classified reference 
set of the same type. 
In practice, this means that a reference set is assembled using expert 
knowledge or a manual classification scheme, which is then used to classify the rest of the 
data using a similarity measure. 
Here, a simple euclidean metric is employed in the 7-dimensional PCA reduced dataset. 
A reference set was assembled using by-eye detections, 
noting the temporal and spatial location in the observations. 
At present, a reference set corresponding to a total number of 958 positions in time and space was assembled. 
This reference set is further divided into 168 PMJ positions (55 separate events) and 790 
background positions of large diversity. These specific numbers were found to yield robust results following  
a trial-and-error approach, studying the number and variability of reference events needed 
until the results were satisfactory and consistent. As the reference set is polled for each automatic identification, 
an unnecessarily large reference set is to be avoided.
 
\begin{figure*}[!tb]
\includegraphics[width=\columnwidth]{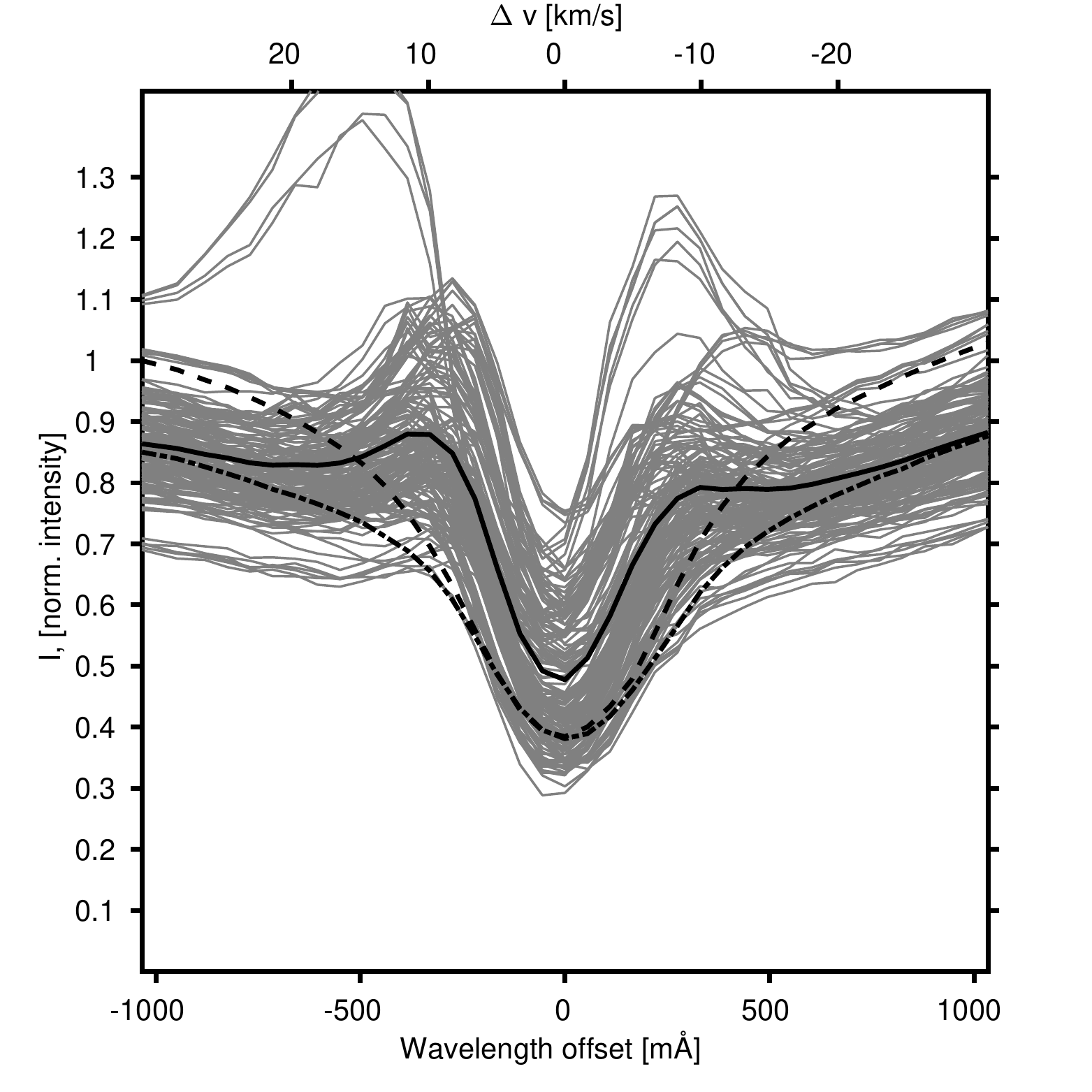} 
\includegraphics[width=\columnwidth]{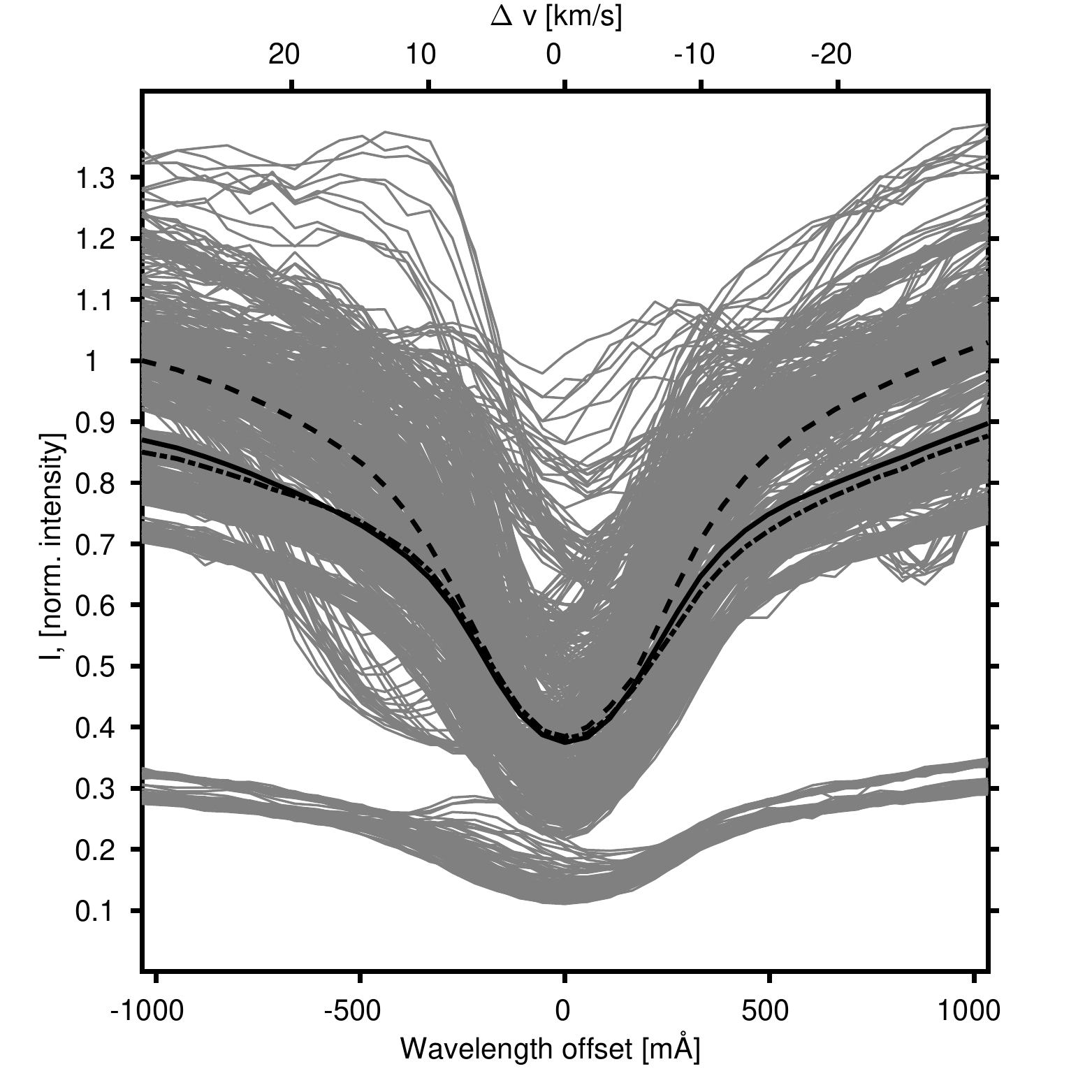} \\
\caption{\label{fig:library_profiles} Line profiles in the \efft line in the kNN reference set used in the automated detections of PMJs. \textit{(Left)} All 168 \efft line profiles 
of by-eye selected positions of PMJs used in the kNN detection scheme (gray-solid) plotted together with their average (black-solid). \textit{(Right)} All 790 line 
profiles of non-PMJ reference positions plotted together with their average (black-solid). For reference in both panels, the averages of the \efft line profile over the whole time series are given for the upper left corner of the full-FOV \textit{(grey-dashed)} and the penumbra \textit{(grey-dot-dashed)} (see Fig.~\ref{fig:sunspot_ffov} for outlines of both areas).}
\end{figure*}

The background positions correspond to datapoints that are clearly not PMJs. 
The k-NN algorithm uses both object and non-object entries in the reference set for identification - for identification each vector in the PCA 
reduced data has its associated distance computed to all points in the reference set. 
A number of k-Nearest Neighbours are then polled, and the point is classified by majority vote.
The selection of the parameter k is performed using an accuracy test of cross-classification of 
the assembled reference set. The classifications resulted in binary maps for each timeframe, 
consisting of background and PMJ detections which were then further processed, 
as outlined in Sect.~\ref{sec:stat}. Figure \ref{fig:library_profiles} shows the \efft line profiles of the 
individual PMJs and background, or non-PMJ positions, that were assembled for the kNN reference set. 
It bears remarking that these profiles are separated for the sake of clarity for the two plots, but form one reference 
set in practice, with profiles marked as PMJ or non-PMJ in the set when polled. The selected PMJ 
profiles have quite the span in terms of intensity throughout the line, but are well-defined by their shape. This is also 
illustrated by their included average, which presents itself as a very well defined PMJ-like profile (which will be made clear in Sect.
\ref{sec:results}). The non-PMJ profiles 
span a wide variety of profiles, sampling positions in the quiet Sun, penumbra, umbra as well as features such as the strong 
inverse Evershed flow present in the observations.

\subsection{Object tracking, statistical analysis and extraction of the \efft\ line-profile  \label{sec:stat}}

The post-processing of the binary maps of raw PMJ detections output by the k-NN 
algorithm was performed in several steps, which are described below. 
First, the binary maps were cleaned of noise and objects tracked through time, see 
Sect.~\ref{sec:cleaning_and_tracking}. 
Following this, base statistics were extracted from the basis of these tracked objects, see Sect.~\ref{sec:statistics_extraction}. 
Lastly, line-profiles in the \efft\ line were found, and this process is detailed in 
Sect.~\ref{sec:computing_profile}.

\subsubsection{Noise removal and object tracking \label{sec:cleaning_and_tracking}}
To remove spurious signals, the binary detections output from the k-NN algorithm were first run 
through an 8-pixel connectivity mask, removing lone falsely identified pixels.
 
Further, a lower ``hard'' area limit of 20 pixels was implemented on object sizes. Additionally, a ``soft'' area limit of 
50 pixels was also implemented. This means that any areas in the original bitmaps below a size of 20 pixels total were not considered at all. The soft 
limit meant that when during tracking no area of a given tracked object reached a size of 50 pixels, the tracked areas were 
discarded. This means that tracked objects could have sizes of below 50 pixels, but only as long as the given object reached a size as large, or larger, than 50 pixels in 
at least one timeframe. This was meant to ensure that it was possible to catch the smaller beginnings of objects, but not track spurious signals 
that never developed and to avoid noise from being tracked for long periods. 

These area limits would correspond to cubic areas with sides $4.47$ pixels $= 194$~km and $7.07$ pixels $= 307$~km for the hard and soft limit respectively, though since a fitting of 
ellipse-shapes is later performed (see below) the spatial limits along one axis may be smaller. 
The limits given may seem prohibitively large, but are chosen after trial and error and yield reasonable results. The final limits were chosen since limits with lower values allowed small-scale noise in the 
bitmaps to be tracked and erroneously labeled as PMJs. This noise was mostly caused by a prominent inverse Evershed flow (see Fig.~\ref{fig:sunspot_ffov}) in the lower right of the observations. 

The object tracking was performed using relatively simple measures of spatial distances in the observations 
to relate discrete detections across timeframes. Here, a maximum distance between center-of-mass pixels of $15$ pixels $= 652$~km between timeframes was used. This is 
equivalent to a projected speed of approximately $52$~km~$^{-1}$, though it must be kept in mind that due to the morphing of detection areas (and thus shifts of center-of-mass pixel 
positions) between frames this threshold can not be seen as a direct threshold on real movement of detected PMJs.
Detections were also restricted to an outline of 
the photospheric penumbra, as seen in the wings of the \efft\ line, see Fig.~\ref{fig:sunspot_ffov}.
Only objects with center-of-mass pixels inside this outline were included in the subsequent analysis. 
Further, spurious detections inside an outline of the umbra were also discarded, also see Fig.~\ref{fig:sunspot_ffov} (these events were most likely related to umbral flashes, 
as was evident when investigating using CRISPEX).

\subsubsection{Basic statistics extraction \label{sec:statistics_extraction}}

A statistical analysis on the resultant final PMJ detections was then carried out.
For estimations of detected object sizes, a fitting of ellipses on the detected objects was carried out. 
The reasoning behind this was the assumption that jet-like objects like PMJs should in principle 
exhibit a somewhat elongated shape. The tracking was thus restricted to PMJs corresponding to 
best-fit-ellipses with an eccentricity of $0.9$, corresponding of an approximate ratio
of semi-major to semi-minor axis in the ellipses of 2.3:1. 
The associated major and minor axes of these fitted ellipses for each single 
object were employed as estimates for the lengths and widths of the given detections respectively (see Fig.~\ref{fig:detections} for 
examples of these fitted ellipses). This allowed for the collection of size-statistics. 
Angular positions around the center of the sunspot for detected objects were also computed. 

\subsubsection{Computing the line profile \label{sec:computing_profile}}

The ``master'' average line-profile for PMJs in the \efft\ line 
was computed using the line-profiles of the center-of-mass-pixels and their 
8 neighbour-pixels for each individual detection. The average was thus carried 
out for only those PMJ detection-areas through all timeframes that actually 
contained the center-of-mass-pixel, as computed from the detection area for 
each detection, as well as this pixel's 8 pixel-neighbours. This selection was 
performed to ensure that only line-profiles of pixel-positions that did not fall 
outside of the actual PMJ detection areas (as could be the case for some 
center-of-mass-pixels due to for example peanut-shaped PMJ areas), and to make 
sure only pixel-positions that represented well-formed PMJs were included. 
The final average line profile is thus based on a subset of the 4253 PMJ 
detections through all timeframes totaling 3953 PMJ detections, as not all PMJ 
areas contained their theoretically computed center-of-mass pixel, or because the 
area did not contain all 8 immediate pixel neighbours of the center-of-mass pixel.

For the computation of the average line-profile a total of 
$3953 \cdot 9  = 35577$ pixel-position line profiles was therefore used. 
The entirety of pixel positions of all PMJ detection areas were 
not employed  for simplicity, and as not to skew the line profiles towards 
larger PMJs, which would then have contributed more strongly towards the 
profile (though a difference in line-profiles between large and small PMJs has 
not been investigated). 

Finally, the individual 9-pixel-average line-profiles 
of each of the PMJs in the mentioned subset were also investigated with regards to 
distinct peaks in the blue and red wing as well as their line-core minimums. 
The wavelength positions of the peaks and the minimum and their final average 
values were estimated by interpolating the individual profiles, and then using
a sliding window approach to find the local peaks and minimum in each given profile. 
Profiles for the average of different categories of profiles (profiles with blue 
and/or red peaks present, profiles with both blue and red peaks present, 
profiles with just either of the peaks present) were also computed to compare to 
the main average and each other. The resulting values for the positions of 
the average minimum, blue and red peaks were calculated by averaging the 
found positions in the individual interpolated 
line profiles, and are thus 
not limited to the 37-sampling points (but are therefore subject to greater 
uncertainty).

\section{Results \label{sec:results}}

The automatic detection of PMJs as outlined in the last section made it possible 
to collect a large and statistically significant set of PMJ detections with 
associated properties that could then be investigated.

In the sections below the different derived properties and statistical values associated with 
these events are presented. 

\subsection{Detection summary}

Table \ref{tab:PMJ_final_stats} summarizes the detections performed by the 
automatic detection pipeline, together with the most basic statistics derived 
from them. It is evident that the automatic approach of identifying PMJs using 
their spectral profile yielded a large dataset from which to infer PMJ 
properties. 
With a number of PMJs equaling 453 tracked events, 
derived statistical values will be significant. Each individual frame in the 
detections contains an average of approximately 21 PMJs, and this rate highlights the 
continuous occurrence of PMJs throughout the observations. As each of these detected PMJs 
is tracked through time, but with unique associated properties in each timeframe, the number 
of individual PMJs summed over all timeframes is greater than the 453 tracked events. 
This number is 4253 PMJs present throughout all individual frames 
and corresponds to the 453 
distinct, tracked events (see table \ref{tab:PMJ_final_stats}). However, some 
further selection is performed before some statistics were computed (namely 
line-profile averaging as well as intensity and line-profile feature 
investigations, see Section \ref{sec:computing_profile}).

\begin{table}[ht]
  \begin{center}
\begin{threeparttable}
  \caption{PMJ detection statistics}
\begin{tabular}{l r} 
\hline
\hline \\
Total PMJ detections (over all frames) [count] &  4253 \\  
Tracked PMJ objects [count] & 453   \\
Mean PMJ detections per frame $\left[\frac{\text{\# events}}{\text{frame}}\right]$ & 21 \\  
Mean lifetime,tracked PMJs (all) [s] &   117   \\ 
Mean lifetime, tracked PMJs (\textless\ 8 min)$^{\text{a}}$  [s]  & 90  \\ 
Mean length$^\text{b}$  [km] &  640 		\\ 
Mean width$^\text{b}$ [km] &  210		\\ 
PMJ minimum position$^\text{c}$ [km~s$^{-1}$] & 0.14  \\  
PMJ blue peak offset$^\text{c}$ [km~s$^{-1}$] & $-10.4$  \\
PMJ red peak offset$^\text{c}$ [km~s$^{-1}$] & $10.2$  \\
\hline
\end{tabular}
\begin{tablenotes}
\item[a] N$_{\text{8 min.}} = 437$
\item[b] Based on ellipse fit.  
\item[c] Based on the average of the 9-pixel average interpolated line profiles.  
\end{tablenotes}
\label{tab:PMJ_final_stats}
\end{threeparttable}
\end{center}
\end{table}

\begin{figure*}
\sidecaption
\includegraphics[width=12cm]{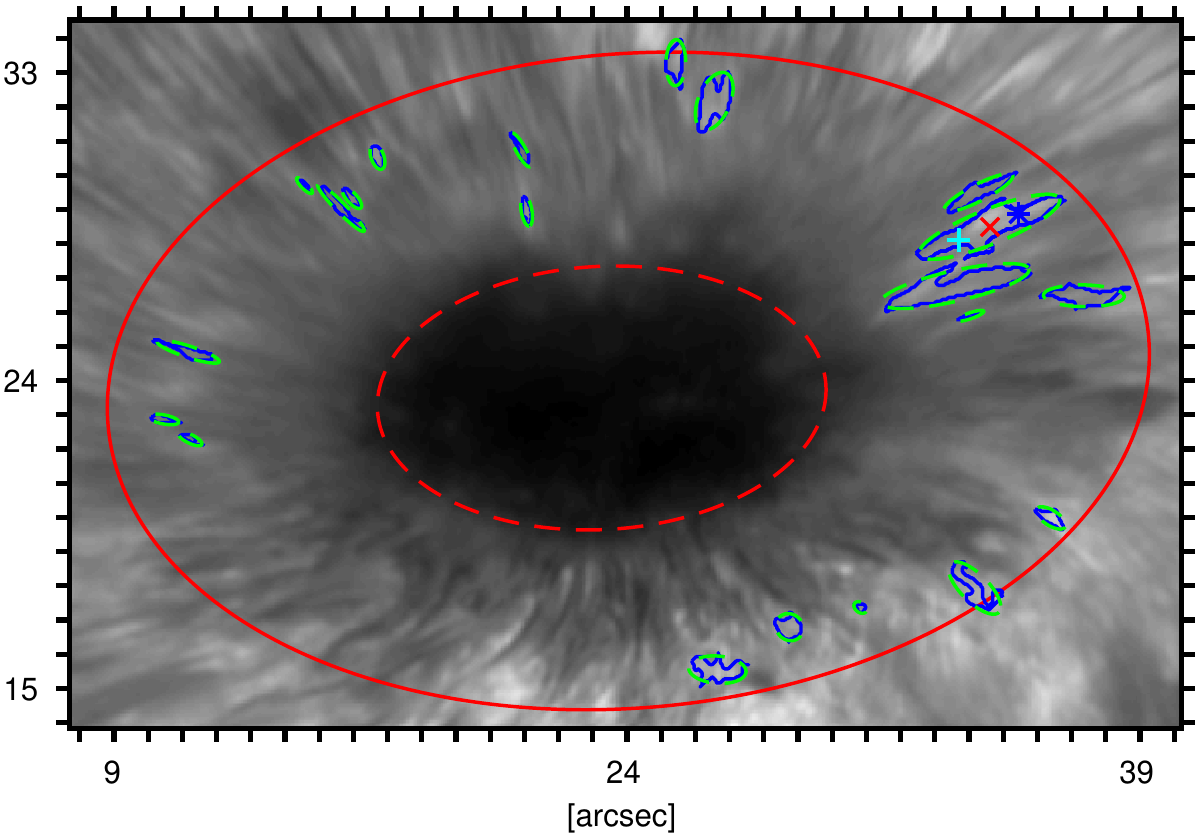}%
\caption{Cropped image in the \efft\ line at an offset of $-275$~m\AA\ with overlain PMJ detection-areas (solid-blue) 
and corresponding computed best-fit ellipses (dashed-green) at time 23 min 27 s. Also indicated for reference are three pixel positions 
inside one of the detection areas, one closer to the sunspot center (turquoise-cross), the center-of-mass-pixel 
(red-x-mark) and one closer to the edge of the penumbra (blue-star). Detailed line profiles for
these pixels are shown in Fig.~\ref{fig:pmj_detail}. \label{fig:detections}}%
\vspace{2ex}\mbox{} 
\end{figure*}

Figure~\ref{fig:detections} depicts a sample frame from the observations at an 
offset of $-275$~m\AA\ 
in the \efft\ line with PMJ detections shown. Visible in the 
figure are the overlain PMJ detection-areas, as well as the associated computed 
ellipses which were used to measure the lengths and widths of the PMJ detections.

As a further reasonability check, a check of the relative brightness in the detection areas of 
all PMJs was performed. Here, the individual average brightness in both the \efft line at an offset of $-275$~m\AA\ and in the 
\ion{Ca}{ii}~H line core was computed for each PMJ detection area. These were normalized to the average brightnesses for each 
individual frame of the penumbral region (as outlined in Fig.~\ref{fig:sunspot_ffov}) in each of the two passbands. PMJs are seen as bright 
features in by-eye detections in both wavelengths, and thus one would expect an average brightening in the detection areas for 
both. Furthermore, as the detections are carried out ultimately
utilizing the \efft line profile, an average brightening in the
detection areas 
in the 
\ion{Ca}{ii}~H line core would further strengthen the co-occurrence of PMJs in both passbands. It was found that $80\%$ of PMJs had an average 
relative brightening above unity in the \ion{Ca}{ii}~H line core and that $79\%$ of PMJs had an average relative brightening above unity in the \efft line 
at an offset of $-275$~m\AA. Furthermore, $71\%$ of detections show a simultaneous brightening above unity in both passbands throughout the 
times series. These values strengthen the assumption that bright features are being detected in the brightness-independent 
detection scheme utilizing the \efft line profile shape. As the brightness of the PMJs was estimated as an average of their entire detection area, 
their brightness as compared to the average of the entire penumbra for a given timeframe may in fact be a conservative comparison. A comparison between 
for example the brightest pixel and the penumbra average for any given PMJ detection would likely yield a higher relative brightness in both passbands. 
This is due to automatically detected PMJs tending to exhibit a larger area than if selected solely by-eye compared to the surrounding intensity. Comparing PMJ brightnesses to the local 
average brightness of the penumbra may very well have resulted in a higher relative brightnesses as well.

\subsection{Basic PMJ statistics and properties \label{sec:basic_stats}}

\begin{figure}[htbp]
\includegraphics[width=9cm]{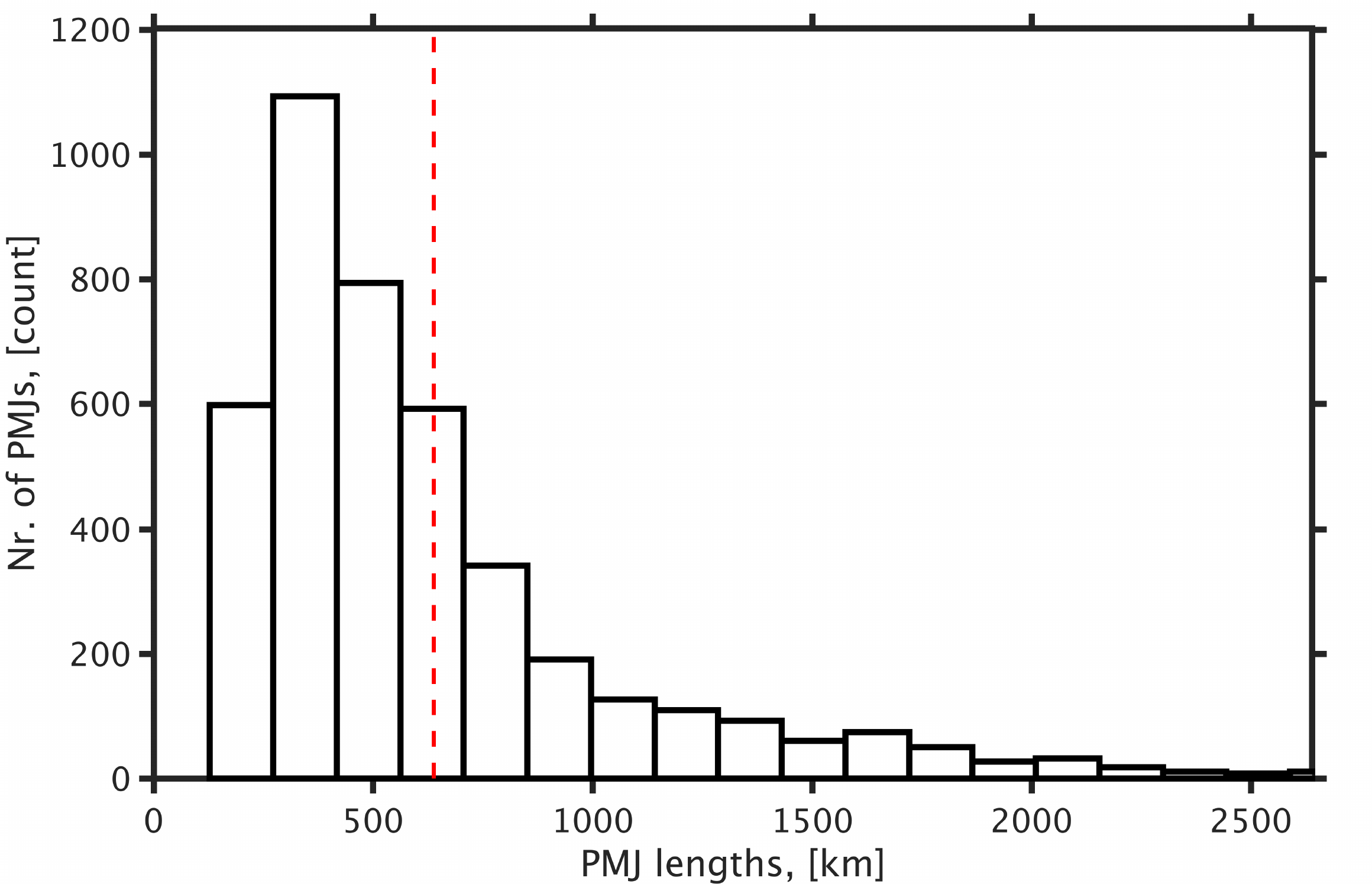} 
\includegraphics[width=9cm]{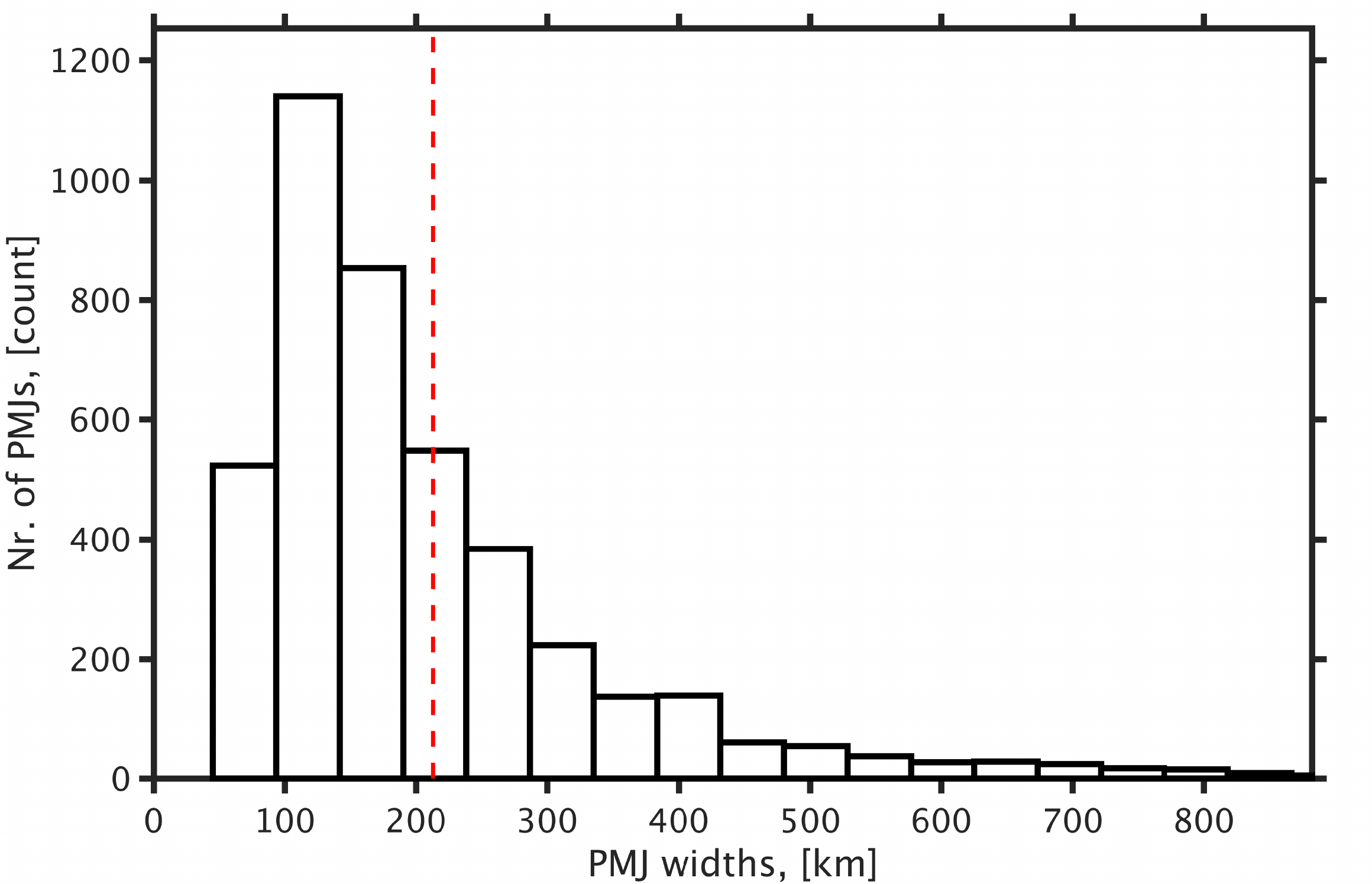} 
\caption{Length and width distributions for PMJs throughout all timeframes. \textit{(Top)} Length distribution with approximate bin size of 145~km, with mean value (dashed) at 640~km indicated, the median value is 489~km. \textit{(Bottom)} Width distribution with approximate bin size of 48~km, with mean value (dashed) at 210~km indicated, the median value is 165~km. The total sample number of individual PMJs is N $= 4253$ \label{fig:lw}}
\end{figure}

Having computed the length and width for all PMJ detections using the 
minor and major axis of the fitted ellipses (as described in Sec. 
\ref{sec:cleaning_and_tracking}), they could be presented as distributions.
Figure~\ref{fig:lw} gives these histogram distributions for the lengths and widths 
of all detected PMJs throughout the observations. 

The distributions for lengths and widths seem to be well behaved with trailing 
ends tending towards zero, as would be expected. The lower ends of the distributions 
also seem to taper off in the last lower length/width bins. One pixel in the 
observations corresponds to 43~km, 
thus both distributions' lowest length/width 
bins with entries include sizes larger than this, however the detections were limited by 
a lower-limit area, which will also effect the lower range. The average 
width/lengths must thus be interpreted with this in mind. The associated mean length of 
$640$~km and mean width of $210$~km (see also Table \ref{tab:PMJ_final_stats}) 
are reasonable, and are consistent in rough magnitude with values reported in 
\citeads{2007Sci...318.1594K}, 
namely lengths of $1000-4000$~km and widths of approximately 
$400$~km or less for the PMJs in Hinode \ion{Ca}{ii}~H. 
These values are also consistent with the mega meter range as 
given in 
\citeads{2013ApJ...779..143R} 
when observing individual PMJs (or penumbral transients) in the \efft\
line. 
Both estimates found here are however still considerably smaller, almost half 
for both values.

\begin{figure}[htbp]
\includegraphics[width=9cm]{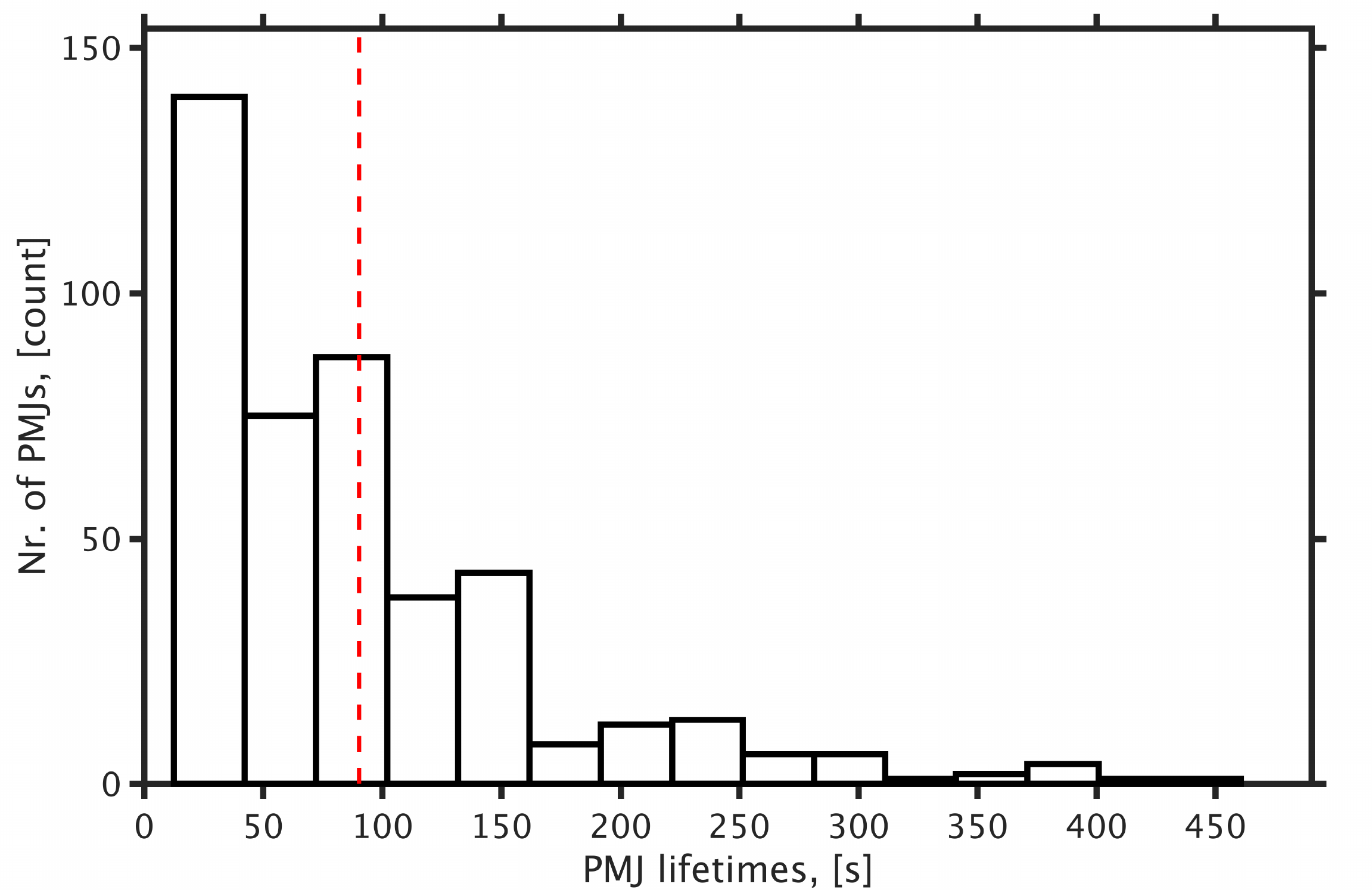} 
\caption{Distribution of tracked PMJ lifetimes with an upper cutoff value of 480~s (8 min) and bin size of 29.9~s. $\text{N}_{\text{8 min. distr.}}~=~437 \quad (96\% \text{ of total tracked PMJs})$. The mean value of 90~s is indicated (red-dashed), the median value is 75~s. \label{fig:dur}}
\end{figure}

In Fig.~\ref{fig:dur}  
the distributions of durations of tracked PMJs is given
with a cut-off value of 8 min, representing $96\%$ of tracked detections, 
not showing outliers. 

These outliers were excluded because they for the most part
represented a small fraction of tracked PMJs that had extremely long 
lifetimes. The truncated distribution represents $96\%$ of tracked 
PMJs. The abnormally long lifetimes were most likely due to detections in areas in which PMJ 
detections were ubiquitous throughout the observations, and where PMJ events 
overlapped so closely in time and space that they were tracked continuously as one event.
Also, spurious and long lasting detections may also have been caused by the strong inverse Evershed flow 
(see Fig. \ref{fig:sunspot_ffov}) present on the center side of the observed sunspot.

\subsection{\ion{Ca}{ii}~8542~\AA\ PMJ line profile \label{sec:profile}}

\begin{figure}[!t]
\centering
\begin{minipage}{\linewidth}%
\centering
\includegraphics[width=\linewidth]{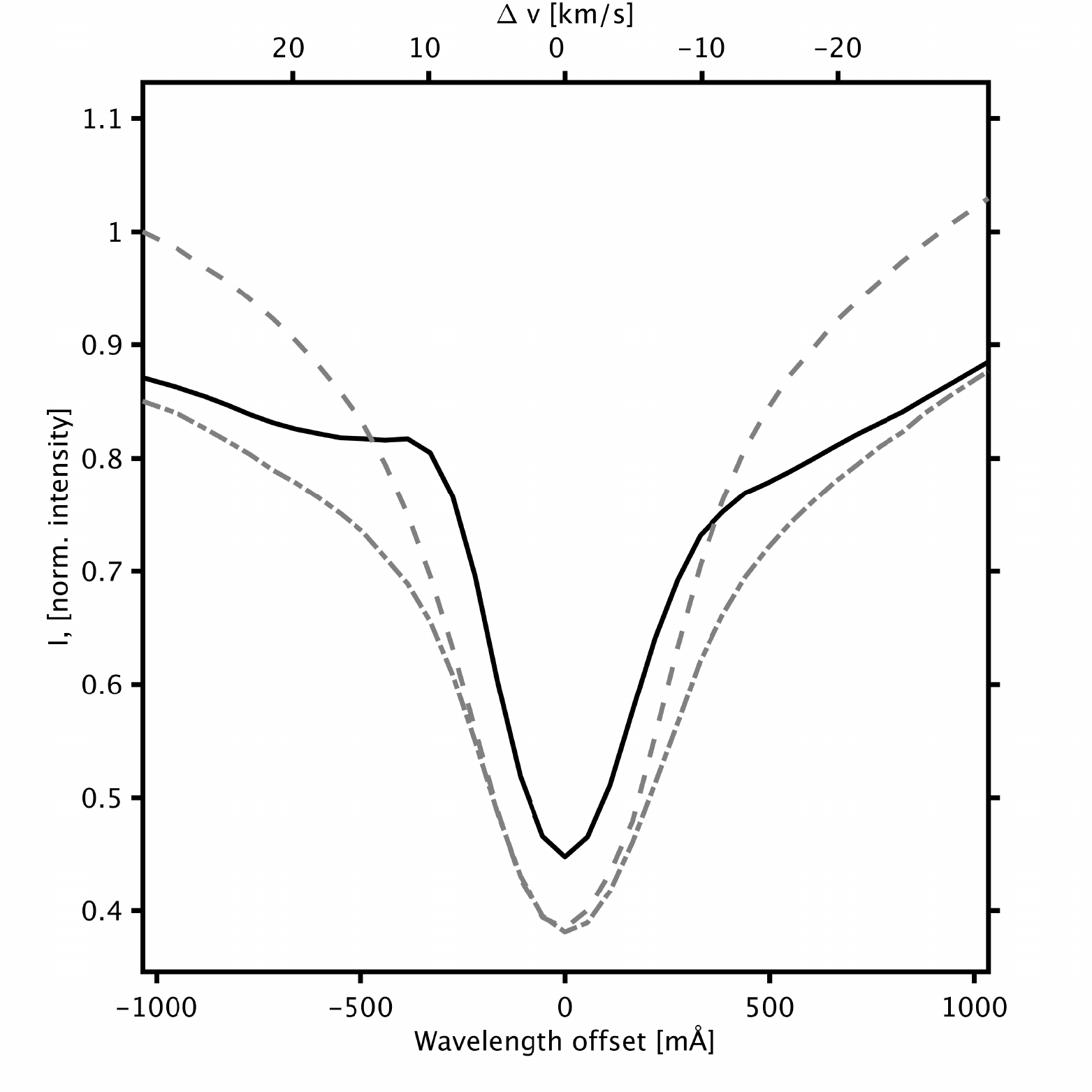}%
\end{minipage} %
\centering
\caption{Averaged \efft line profile of PMJ center-of-mass pixels and their 8 neighbour-pixels \textit{(black-solid)}. For reference, the averages of the \efft line profile over the whole time series are given for the upper left corner of the full-FOV \textit{(grey-dashed)} and the penumbra \textit{(grey-dot-dashed)} (see Fig.~\ref{fig:sunspot_ffov} for outlines of both areas). \label{fig:pmj_profile}}%
\end{figure}

The PMJ line-profile in the \efft\ line computed from the average of the $3953 \cdot 9  = 35577$ 
pixel-positions is given in Fig.~\ref{fig:pmj_profile} together with 
reference profiles for the quiet Sun and the penumbra. 
The PMJ profile is characterized by enhanced inner wings (at about
$\pm385$~m\AA\ and a brighter line core (with the core being at 116\% of the quiet sun line core brightness).
The enhancement of the inner blue wing is stronger than in the red
wing. 
The far wings of the PMJ profile approach the level of the average
penumbra. 
This average profile is less pronounced than individual profiles of PMJs selected by eye and 
inspecting individual pixel-position profiles (which will be made more evident in section \ref{sec:conti_case}).

The found average PMJ profile is however still consistent with other reported profiles as given in
\citeads{2013ApJ...779..143R} (their Fig.4) 
and 
\citeads{2015ApJ...811L..33V} (their Fig.5), 
but does show a less pronounced peak in the blue, and a very weak enhancement in the red 
compared to most of these published profiles, but is still
recognizable. 
It must be emphasized that the profile presented here is an average, computed 
from many individual profiles, whilst the profiles it has been compared to in 
previous work are profiles from individual pixel positions of by-eye-selected 
PMJs. These individual profiles were most likely selected specifically 
due to their distinct features, meaning that the average profile presented 
consequently will not present as sharp features. Individual profiles of PMJs in 
the presented observations (both selected by eye, as well as those contained in 
PMJ detection areas) still exhibit such distinct features to a large degree, 
as will also be exemplified in Sec. \ref{sec:conti_case}.

\begin{figure}[!t]
\centering
\begin{minipage}{\linewidth}%
\centering
\includegraphics[width=\linewidth]{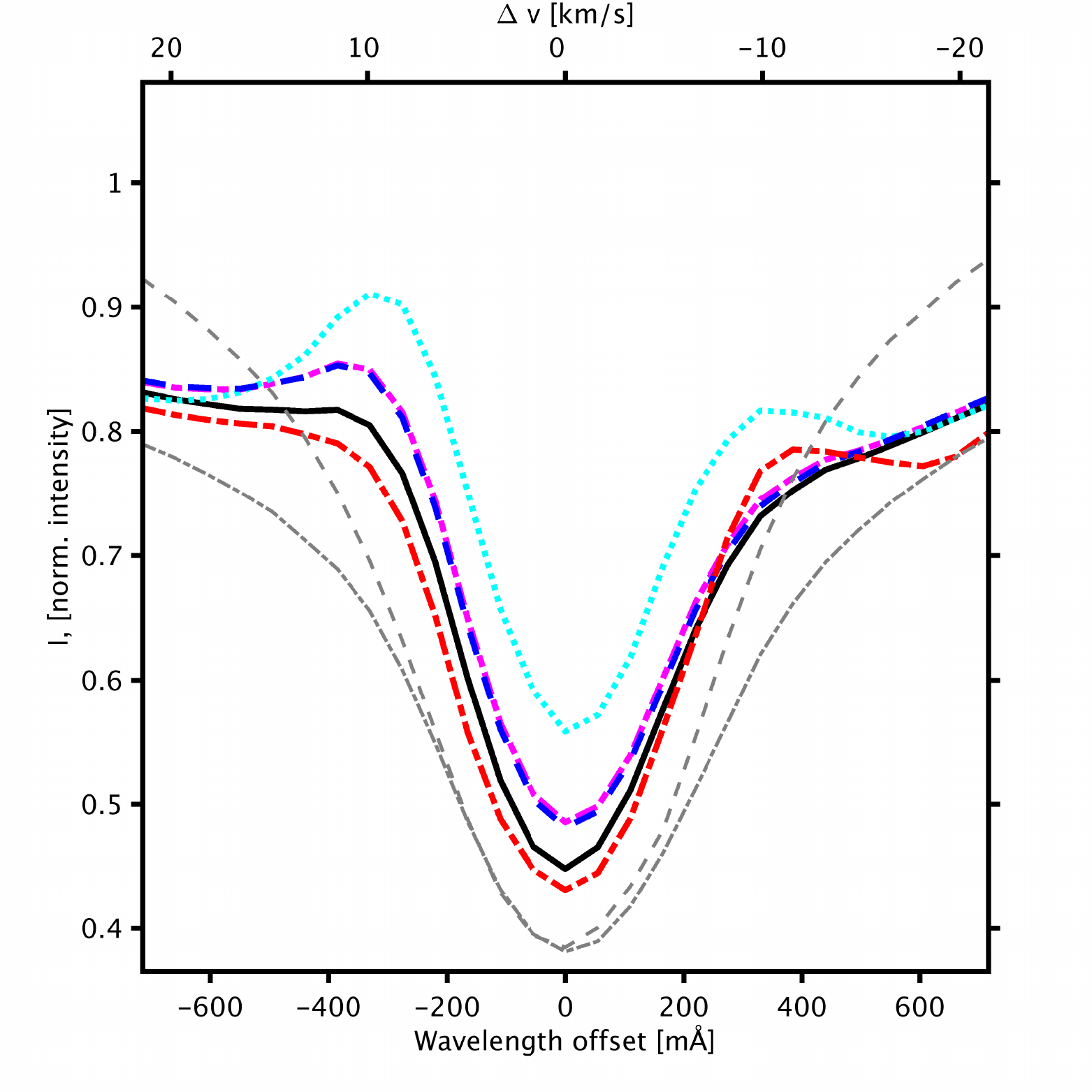}%
\end{minipage} %
\centering
\caption{Average \efft PMJ line profiles divided by the presence of clear
  inner wing peaks: 
PMJs with only distinct blue peaks (N$_{\text{only blue}} = 1637$, \textit{blue-dashed}),
PMJs with only distinct red peaks (N$_{\text{only red}} = 29$, \textit{red-dashed-dotted}), 
PMJs with both distinct blue and red peak simultaneously (N$_{\text{blue \textit{and} red}} = 114$, \textit{cyan-dotted}),
and PMJs with distinct blue and possible red peaks (N$_{\text{blue \textit{with possible} } red} = 1725$, \textit{magenta-dashed-dotted}, note that this profile nearly
overlaps with the only-distinct-blue-peak profile).
Included for reference are the profiles of Fig.~\ref{fig:pmj_profile}: 
all PMJs used in the line-profile averaging (\textit{black-solid}), 
average penumbra \textit{(grey-dot-dashed)} and quiet Sun
\textit{(grey-dashed)}. 
}
\label{fig:pmj_profile_with_peaks}
\end{figure}

As mentioned in section \ref{sec:computing_profile}, when averaging the master line profile, 
each individual (9-pixel average) profile corresponding to individual detections was also inspected 
for the presence of a tell-tale blue and/or red peak in the line. 
As a result, it was possible to create averages of the profiles according to the presence of peaks in the blue or red. Thus, averages for 
the profiles for the cases of both a blue \textit{and} red peak being present, one \textit{or} the other being present or just one of the 
peaks being present could be computed. Figure~\ref{fig:pmj_profile_with_peaks} shows these average \efft line profiles for different subsets 
of PMJ detections. 

For completeness, Table~\ref{tab:PMJ_peak_stats} summarizes these different groups of profiles. Note especially that only 
less than half (N$_{\text{profiles with peaks}} = 1868$) of all 9-pixel average PMJ profiles used in the peak detection (N$_{\text{all}} = 3953$) have clear automatically detectable peaks.
Further, a vast majority 
of these were in the blue, with a total number of blue peaks of N$_{\text{blue peaks}} = 1725$ and a total of red peaks of N$_{\text{red peaks}} = 143$. This 
corresponds to percentages of $92\%$ and $8\%$, of blue and red peaks respectively, of the total number of detected peaks.
Further, of all investigated averaged 9-pixel profiles, $43\%$ exhibit blue peaks whereas only $3.6\%$ exhibit detectable red peaks.

\begin{table}[!t]
  \begin{center}
  \caption{Summary of PMJ peak-presence in individual profiles}
  \begin{tabular}{l l} 
\hline
Nr. of (9-pixel average) profiles &  \\
used in peak detection & 3953\\
Nr. of peaks detected & 1868\\
\hline
Peak combinations  & Nr. of profiles   \\
in profiles & \\
\hline
Blue \textit{and/or} red peak & 1780	\\ 
Both blue \textit{and} red peak  & 114 \\
\\
Blue peak (with possible red) & 1725 \\
Only blue peak & 1637 \\
\\
Red peak (with possible blue) & 143 \\
Only red peak &	29  \\
\hline
\end{tabular}
\center{}
\label{tab:PMJ_peak_stats}
\end{center}
\end{table}

From Fig.~\ref{fig:pmj_profile_with_peaks} 
it is evident that the average line profile of those profiles that have either both or one peak present (magenta-dashed-dotted in the figure) is near 
identical to the average line-profile for the profiles that have blue
peaks (and no red peaks) present (blue-dashed in the figure). 
This is concurrent with the fact that there 
is a much larger number of profiles with blue peaks compared to red
peaks. 
The average line-profile of profiles with only red peaks is correspondingly significantly different from both these 
profiles and only shows a peak in the red (note however the low number of 29 profiles in this average). 
All three profiles are more distinct in appearance than the overall average line-profile of all used (9-pixel) line profiles (solid-black in 
the figure). 

Finally, and most notably, the average line-profile of the profiles with both blue and red peaks present simultaneously, which therefore has well-defined peaks in 
both the blue and red, more strongly resembles the profiles of by-eye selected PMJs in the present dataset (which again, will be made more evident in section \ref{sec:conti_case}). 
For likely the same reason, this profile also more strongly resembles reported profiles 
as found for by-eye detected PMJs in the \efft line given in 
\citeads{2013ApJ...779..143R} (Fig.4)  
as well as several of the distinct profiles presented in  
\citeads{2015ApJ...811L..33V} (Fig.5).  
This profile is also of overall greater intensity in both the peaks as well as in the line-core minimum compared to the overall PMJ average 
profile and the FOV average of the \efft line. In fact, the line core is at an 145\% intensity compared to the quiet sun line core average, a clearly
greater enhancement than for the overall line profile average, which as given earlier exhibited an intensity of only 116\% compared to the quiet sun. 
This makes it plausible that the higher overall intensity in these types of PMJs makes 
it easier to pick them out by eye, and thus makes the presence of both peaks in these selections more likely, whereas the automatic 
detection presented may not be as susceptible to this bias (after the initial selection of the k-NN reference set). 
An over-selection of PMJs with strong enhancements in the red wing of the \efft line may therefore be likely in by-eye detections.

It is worth noting that the average PMJ profile of the kNN reference set shown in Fig. \ref{fig:library_profiles} exhibits a similar shape 
to the average profile of detected PMJs with both blue and red peaks present, and thus has more clear peaks than the final PMJ average profile of all 
PMJ detections. It also has an overall higher intensity than the total average, though not as high as the average with both blue and red peaks 
present. This similarity is most likely due to the previous point raised that by-eye selection favours PMJs with strong enhancements in 
both the blue and the red, as well as overall brighter PMJs.
%

\begin{figure}[!t]
\includegraphics[width=8.0cm]{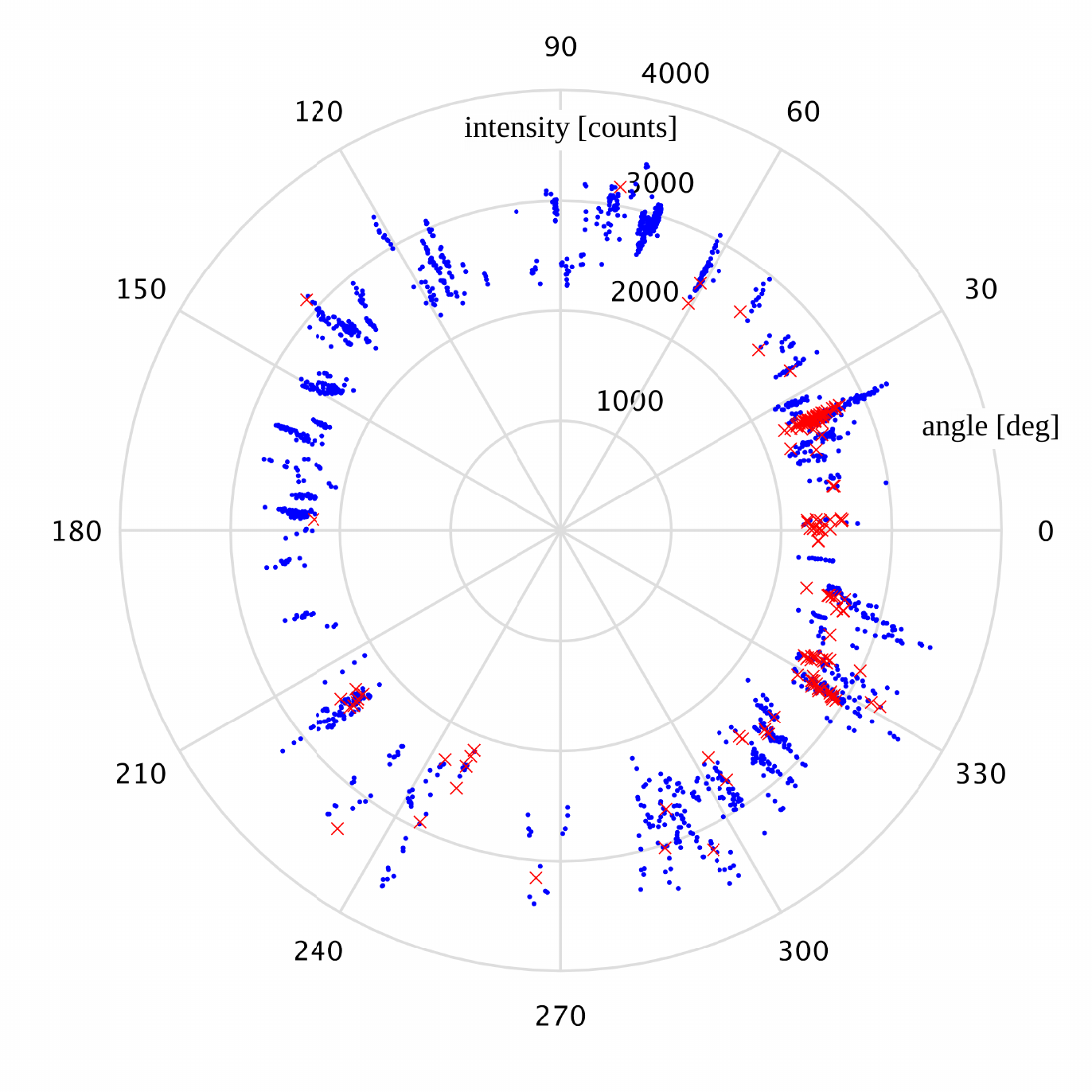}%
\centering%
\caption{Angular position inside the penumbra (and centered on the
  umbra) along the polar axis [degrees], vs. peak intensity [counts]
  of blue and red peaks in PMJ profiles in the \efft line. Plotted are
  the peak-intensities in the respective wing for PMJs with
  automatically identifiable peaks in the red (crosses, $N$=143) and blue
  (dots, $N$=1725) wings of the line. The direction to disk center is at
  approximately 270$\degree$. \label{fig:angle_intensity}}%
\end{figure}

We investigate the spatial distribution of the PMJ profiles with
distinct blue and red peaks. 
It may be well plausible that the average PMJ orientation is related to the
large scale magnetic field topology of the sunspot so that the
inclined viewing angle may have an effect on the spatial distribution
of the observed PMJ profiles. The results are presented in
distribution graphs in polar coordinates, centered on the umbra and
the limb direction approximately $90\degree$.

In Fig.~\ref{fig:angle_intensity}, 
a scatter plot of the peak-intensities versus the angle around the center of the sunspot of the automatically detected peaks in the red and blue of the \efft line for the detected PMJs is given. 
There is no clear discernible bias in the plot with regard to the intensity, both for the blue and red peaks. However, there is a readily apparent clustering of red peaks in 
the degree range of $315 - 30$, with two noticeable groups within this range. The group above the zero-degree mark seems to neatly coincide with the principal PMJ hot-spot 
as described in Sec. \ref{sec:conti_case} below (see Fig.~\ref{fig:density}). The presence of more readily detected red peaks in an area in which many PMJs are found throughout 
the timeseries is intuitive, as a high count of PMJs should lead to a higher count of red peaks as well. On the other hand, the area still seems overrepresented in the amount of 
red peaks compared to other areas with significant numbers of PMJs, and may indicate that red peaks are favored in some areas.

\begin{figure}[!t]
\includegraphics[width=8.0cm]{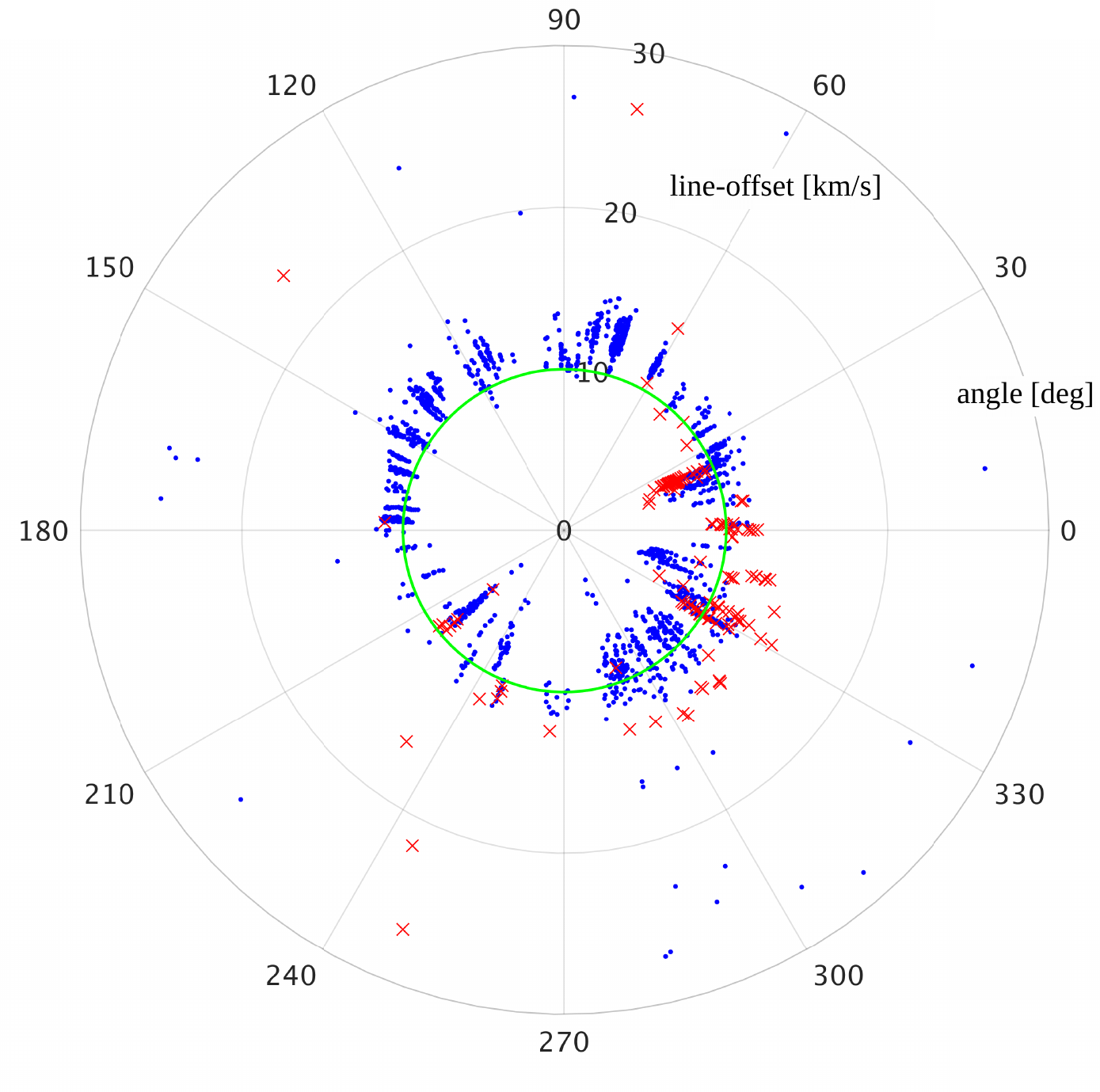}%
\centering%
\caption{Angular position inside the penumbra along the polar axis
  [degrees], vs. both absolute blue- and red-shift from
  \efft line-center along the radial axis [km~s$^{-1}$]. Plotted are offsets
  for PMJs with automatically identifiable peaks in the red (red-crosses)
  and blue (blue-dots) wings of the line. Also marked is a reference at
   $10$~km~s$^{-1}$ (green-solid). 
\label{fig:angle_line_shift}}%
\end{figure}

In Fig.~\ref{fig:angle_line_shift}, 
a scatter plot of the wavelength-positions versus the angle around the center of the sunspot of the automatically detected peaks in the red and blue of the \efft line for the PMJ detections is given. Here, there is an 
apparent bias in the degree of red shift in some of the red peaks, whereas there is no clear bias in the blue shift of the blue peaks. We can again discern the general clustering of red peaks in the same range as for Fig.~\ref{fig:angle_intensity}, but 
this time there is a clear difference in the two groups within this
angular range. The group coinciding 
with the PMJ hot-spot (see again Sec. \ref{sec:conti_case} and Fig.~\ref{fig:density}) at slightly below 30 degrees has a clearly lower 
red-shift, and is less spread out in values, than the group of red peaks clustered around 330 degrees (see also the reference line at $10$~km~s$^{-1}$ in the Figure). The atypical red shift of the red peaks situated around the 330 degree mark are likely largely 
caused by a strong inverse Evershed flow that moves into the penumbra
at this location, which may also account for the greater overall
spread of red shifts in this region.
This region of strong inverse Evershed flow is an extension of the
large dark cloud in the bottom middle part of the right panel of
Fig.~\ref{fig:sunspot_ffov}. 
We observe these inverse Evershed ``clouds'' move into this part of
the penumbra throughout the full duration of the time series. 
A detailed inspection of the spectral profiles with CRISPEX of this
region reveals that at times, the profiles are largely affected by
this inverse Evershed flow.

There also seems to be an overall higher blueshift in the detected blue peaks 
of the \efft line profiles on the limb-side of the Penumbra (in the 30 to 180 degree range), which is 
counter intuitive to a line-of-sight enhancement of the blueshift, as this would be expected for the 
disk-side instead. This will be mentioned further in Section \ref{sec:disc}.

\subsection{Clustering and a near-continuous occurrence of PMJs \label{sec:conti_case}}

\begin{figure*}
\sidecaption
\includegraphics[width=12cm]{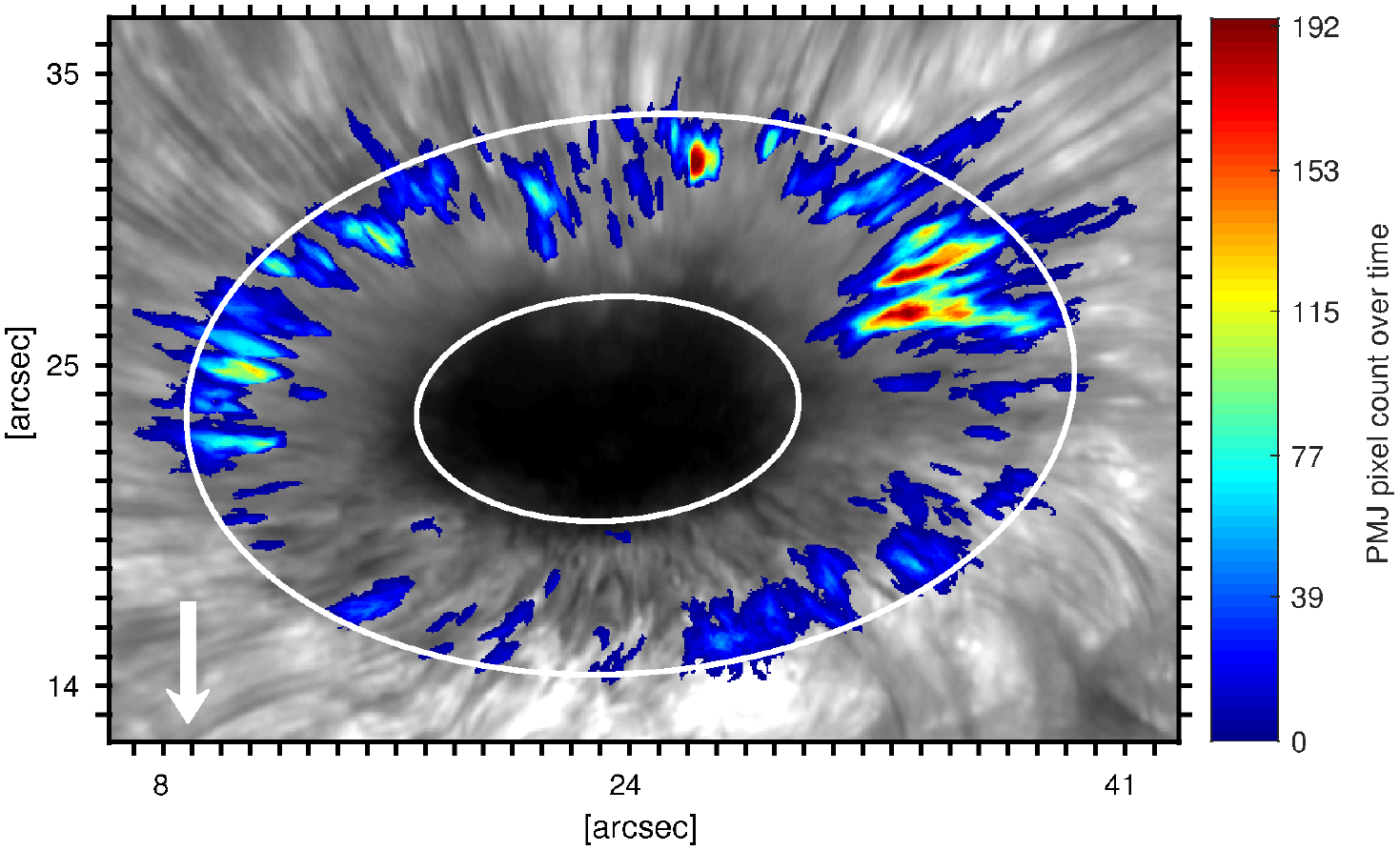}%
\caption{PMJ densities, with all pixel-detections 
 summed over all 202 timeframes
, overlain a frame at the midpoint in time of the observations, at an
offset of $-275$~m\AA\ 
in the \efft line. The arrow indicates the direction towards disk-center. \label{fig:density}%
\vspace{2ex}\mbox{} 
}
\end{figure*}

Figure~\ref{fig:density} depicts a density map of all the individual PMJ detection-pixels, summed over all timeframes.
PMJs are detected rather evenly distributed over the azimuthal
direction, and mostly in the outer half of the penumbra. 

From the density distribution it is evident that 
there are preferred sites for PMJ formation and that they are not evenly distributed throughout the sunspot's penumbra. We can see a clear clustering 
of PMJ detections in certain regions, and two distinct regions in the upper right corner in particular. These regions are the sites of a large number of PMJs 
throughout the observations.  

There is an apparent bias in the number of detections with regards to
position in the penumbra, as it is readily seen that there are many
more detections on the limb-side of the sunspot. The right-upper
corner of the penumbra mostly coincides 
with the 
far-side of the observer's line-of-sight, which might possibly contribute to a larger number of clear detections. Reciprocally, the lower side of he sunspot may exhibit a lesser amount of detections due to foreshortening of the (nominally) 
elongated PMJs. There are however still distinct hotspots of PMJ-activity on the same side of the sunspot, meaning that a clustering of PMJs could not be caused by foreshortening effects to a large degree.

\begin{figure*}[!htbp]
\includegraphics[width=\textwidth]{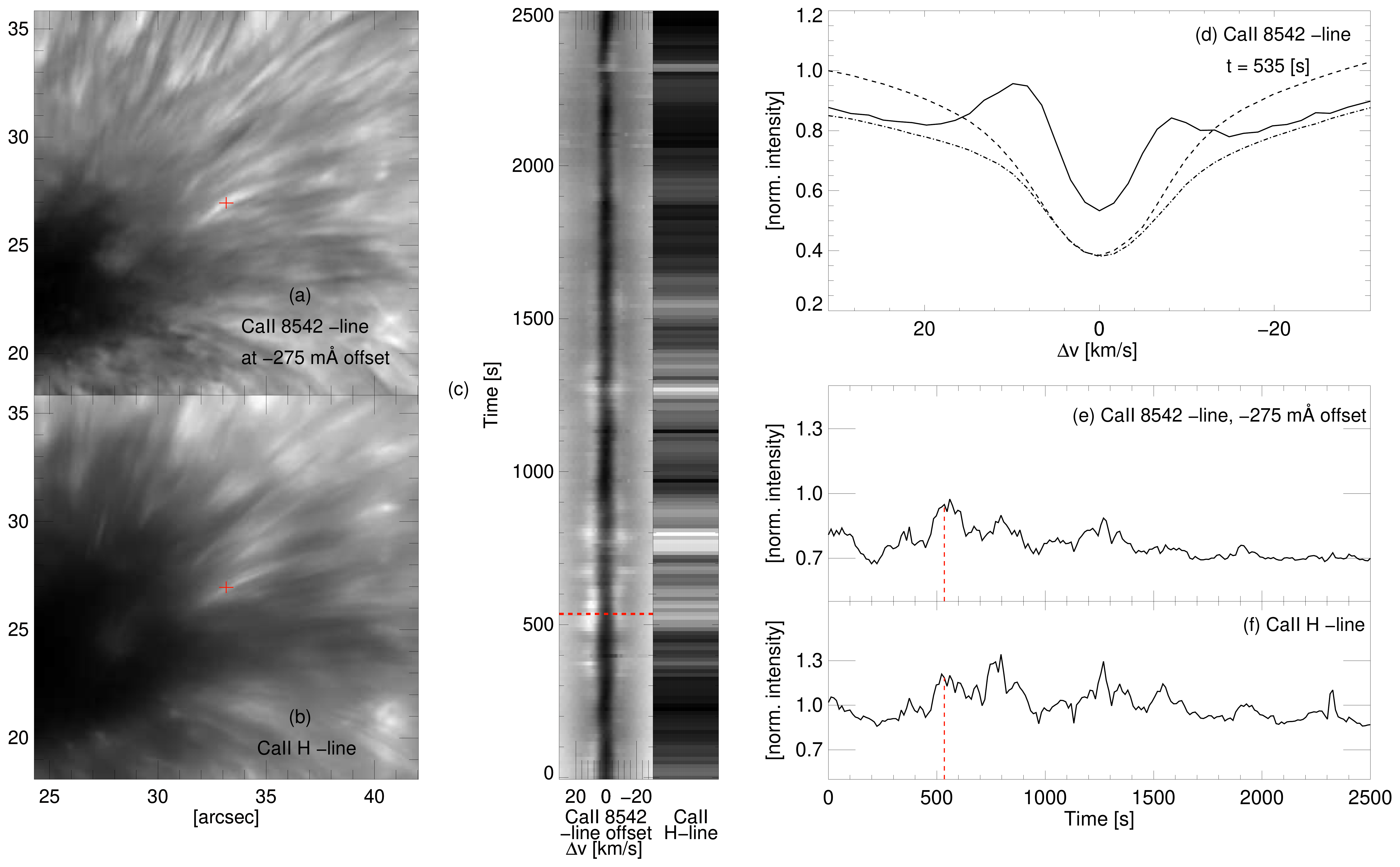} 
\caption{PMJ event studied in detail at time $= 535$ s. (a) PMJ marked by a cross in a subfield of the observations in the \efft line at an offset of -275 m\AA . (b) PMJ marked by a cross in a subfield of the observations in the \cahline. (c) Timeslices for the full duration of the observations in the sampled \efft line (left) with time of the event indicated (dashed line), and in the monochromatic \cahline\ core (right). (d) \efft line profile of the PMJ event (solid line) with the average of the line over the upper left section of the full FOV (dashed) and the average of the line over the penumbra (umbra excluded) (dash-dotted) (see Fig.~\ref{fig:sunspot_ffov} for these regions) with time of the event given. (e) Intensity curve in the \efft line (solid line) at an offset of -275 m\AA\ for the full duration of the observation at the PMJ event location with PMJ event time (dashed line). (f) Intensity curve in the same position and duration in the \cahline\ core (solid line) and event time (dashed).\label{fig:compare_slice1}}
\end{figure*}

Figure~\ref{fig:compare_slice1} highlights the behavior of a 
``PMJ-hotspot'' (a site of repeated PMJ activity) picked out by-eye using CRISPEX (see legend for details). The timeslices in particular, as well as the intensity-time plots, highlight how the PMJ-events  seem to present as continuous processes of waxing and waning in
intensity in both the \efft\- and the \cahline, as opposed to very
well-defined one-off events with clear onsets and ends. Both events
highlighted in these figures are situated in a PMJ-hotspot as readily seen in Fig. \ref{fig:density}.
We can observe the typical behavior of PMJs in the present observations, namely being localized to specific regions, and to seemingly reoccur over time. 
This ongoing process of PMJ generation at preferred sites may possibly be due to favorable magnetic field structures at these particular sites, leading 
to repeated magnetic reconnection in quick succession. 

\begin{figure}[htbp]
\includegraphics[width=\columnwidth]{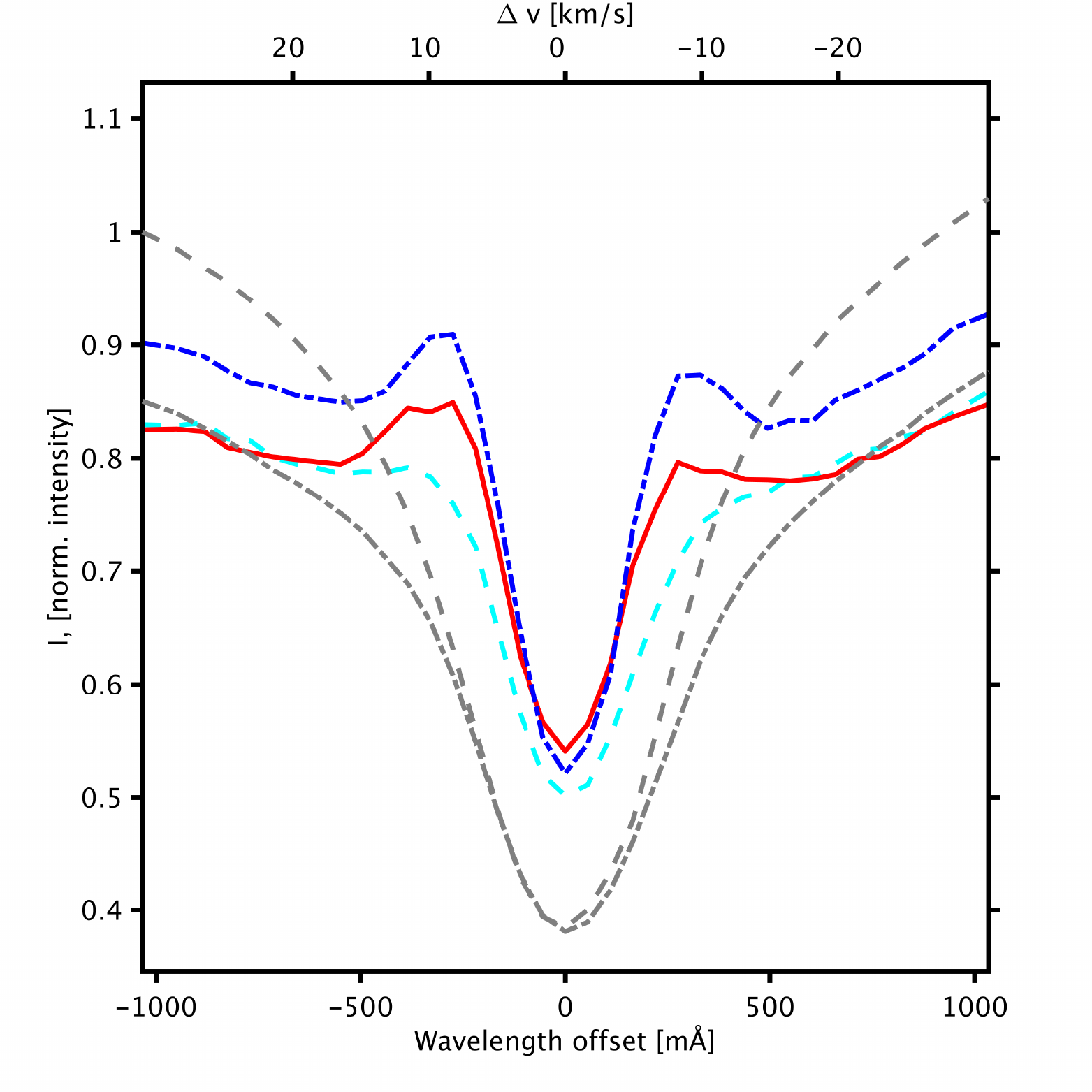} 
\caption{\efft line-profiles normalized to average quiet Sun line profile at different positions within a PMJ event detection area. The corresponding positions and the detection border of the event are marked in Fig.~\ref{fig:detections}. Three specific PMJ line profiles are shown, one  situated closest to the umbra (turquoise-dashed, turquoise-cross in Fig.~\ref{fig:detections}), the center-of-mass pixel position (red-solid, red-x-mark in Fig.~\ref{fig:detections}) and one furthest from the umbra (blue-dot-dashed, blue-star in Fig.~\ref{fig:detections})
Reference averages of the \efft line profile over the whole time series are given for the upper left corner of the full-FOV \textit{(grey-dashed)} and the penumbra \textit{(grey-dot-dashed)} (see Fig.~\ref{fig:sunspot_ffov} for outlines of both areas). \label{fig:pmj_detail}}
\end{figure}

Figure~\ref{fig:pmj_detail} highlights the \efft\ line profiles of three 
specific pixel positions, all contained within a single PMJ detection area, as 
shown in Fig.~\ref{fig:detections}. Shown in Fig.~\ref{fig:pmj_detail} 
are the profiles for a pixel position on the umbra-side of the detection area, 
the center-of-mass pixel position of the area, and a pixel position on the outer side of the area. 
These line profile examples highlight that the automatic detection 
envelops a larger area than what would perhaps be picked out as a singular PMJ 
in by-eye detections (see Fig.~\ref{fig:detections}), but that the contained 
profiles still have the distinct PMJ shape. 
In particular, this event may be categorized as two distinct jets using by-eye 
detections at the shown wavelength offset, whereas the automated 
method based on the line profiles identifies the entire area as one PMJ. The 
three different profiles highlighted all exhibit the distinct PMJ peak in the 
blue. There is also a clear enhancement in the red, but this is most 
evident for the center-of-mass position and the reference point closest to quiet Sun. The 
reference profile for the umbra-side position has some enhancement in the red, but is 
less clear. The two profiles closest to the umbra are very much distinct PMJ profiles, whereas 
the center-side profile is generally more subdued, but still with enhancement in the blue and red 
peak positions. The profile also has as a clearly enhanced line core compared to the penumbra average line profile.
These profiles are generally very similar to by-eye selected profiles (as for example seen in Fig.~\ref{fig:compare_slice1}). 
The profiles in Fig.~\ref{fig:pmj_detail} also exhibit an incremental increase in intensity 
for the blue and red peak wavelength offsets as one moves towards the quiet sun for the overall profiles. This 
trend is not present for the line core however, with the line core intensities seeming fairly stable.

\section{Discussion \label{sec:disc}}    

\subsection{Similarity to Ellerman bombs}

The typical PMJ \ion{Ca}{ii}~8542 spectral profile with enhanced inner
wings resemble the characteristic spectral profile of Ellerman bombs,
a similarity that was already pointed out by
\citeads{2013ApJ...779..143R}  
and
\citeads{2015ApJ...811L..33V}. 
Ellerman bombs
\citepads{1917ApJ....46..298E}  
are the telltale signature of magnetic reconnection in the low
atmosphere, usually associated with emergence of strong magnetic flux
in active regions
(see 
\citeads{2013JPhCS.440a2007R} 
for a recent review). Like for Ellerman bombs, magnetic reconnection is the 
driving mechanism for PMJs that is favored in the literature (see e.g. 
\citeads{2007Sci...318.1594K}, 
\citeads{2008ApJ...686.1404R} 
and 
\citeads{2012ApJ...761...87N}
).
Ellerman bombs are commonly observed in H$\alpha$ where they exhibit
enhanced wings but an undisturbed line core which is an indicator that
the reconnection occurs below the chromospheric canopy 
\citepads[see, e.g., ][]{2011ApJ...736...71W}. 
\citeads{2013ApJ...774...32V} 
studied Ellerman bombs in H$\alpha$ and \ion{Ca}{ii}~8542 from a CRISP
data set that was acquired just before the data we present here. 
The \ion{Ca}{ii}~8542 Ellerman bomb profile has enhanced wings like in
H$\alpha$ (see their Fig.~5) and
a dark line core from obscuration by overlying chromospheric
fibrils. 

It is worth to emphasize that the wing enhancement in PMJs is much
more modest than in Ellerman bombs where the intensity level in the
inner wing rises above the far wing quiet Sun level and reaches far
into the wings, beyond 1~\AA\ offset from the line core. 
In PMJs the enhancement extents mostly to about 0.6~\AA\ and stays well below
the far wing quiet Sun reference intensity. 
Further, we note a 116\% intensity of the \ion{Ca}{ii}~8542 average PMJ line core as compared
to the quiet Sun (and a 145\% intensity for the line profile average with both blue and red peaks present). 
In the line core images, PMJs are not obviously obscured by
overarching fibrils like for Ellerman bombs, and sometimes we can
discern bright features that resemble the PMJ morphology at
$-275$~m\AA\ offset, see the bottom panel of Fig.~\ref{fig:prev_H_22}
for an example. 
We regard the enhanced line core as consistent with the observation that PMJs get heated
to transition region temperatures as reported by 
\citeads{2015ApJ...811L..33V} 
who found signs of progressive heating along the PMJ length from
coordinated IRIS and SST observations. 

\subsection{Spatial distributions}

Besides the enhancement of both inner wings in PMJ line profiles, we
note that there is a preference for larger enhancement of the blue
wing. 
For those profiles where the enhancement is in the form of one or two clearly identifiable inner wing peaks
(close to half of the detected PMJs), 
the majority has a blue peak. 
This is also reflected in the average PMJ profile that displays a
clear blue-over-red asymmetry.
We investigated whether there is any trend in the spatial distribution
of the properties of the inner wing peaks.
If one would (perhaps naively) interpret PMJs as near-vertical plasma upflows, one would
expect this to have an imprint on the observed spectral profiles for a
sunspot under this observing angle (observing angle
$\theta=57\degr$). 
This viewing angle effect is for example very clear for the photospheric Evershed
flow in sunspots, where Dopplermaps show a clear, highly inclined, outflow in the
form of red-shifts at the limb-side and blue-shifts at the center-side
for sunspots away from disc center
\citepads[see, e.g.,][for a recent example]{2011Sci...333..316S} 
We find, however, no clear systematic imprint on the spatial distribution of
spectral parameters of PMJ profiles. 
The peak intensity of the inner wing peaks show no trend in the
spatial distribution over the sunspot (see
Fig.~\ref{fig:angle_intensity}).
There may be a trend of higher blue-shifts of the blue peaks at the
limb-side penumbra (Fig.~\ref{fig:angle_line_shift}) where the blue peaks are all shifted
more than $-10.5$~km~s$^{-1}$.
On the center-side, the shifts are between $-7$ and $-14$~km~s$^{-1}$.
This trend goes against the simple interpretation of the shift of the
blue peak as a pure Dopplershift from a upflow in the expanding
chromospheric magnetic field of the sunspot -- in such scenario the
strongest blueshifts would be found in the center-side penumbra. 
We conclude that detailed numerical modeling with a realistic
treatment of the radiative transfer in the optically thick penumbral
atmosphere is required in order to interpret the PMJ spectral
profiles. 
A complicating factor is the inverse Evershed effect in the form of
(mostly) red-shifted clouds that at certain
times and spatial locations result in strongly affected line
profiles. 
Part of the profiles that were identified as ``red-peak'' PMJ profiles
were clearly affected by inverse Evershed clouds that were unrelated
to PMJs.

\subsection{Spatial dimensions}

From our large statistical sample of automated PMJ detections, we
determine PMJ lengths that are on the 
short side as compared to the 
measurements from Hinode Ca~H observations 
\citepads{2007Sci...318.1594K}. 
To large extent this can be attributed to our method's sensitivity to
weaker events and the inclusion of the center side penumbra, where the
projected PMJ extensions suffer from foreshortening. 
The measurements from previous studies have an intrinsic bias towards
longer PMJs from by-eye selection. 

\subsection{Lifetimes}

We determine an average lifetime of 90~s which is longer as compared to the
typical lifetime of less than 1 min reported earlier
\citepads{2007Sci...318.1594K}. 
However, we note that we find a large number of short-duration events
(the median lifetime is 75~s).
We have discarded long-duration detections ($>$8~min) 
that were in part resulting from clustering of individual PMJs occurring in close vicinity and
rapid succession, and in part due the earlier described strong inverse Evershed flow on the disk side 
of the sunspot, distorting the \efft line profile, causing false identifications. 
Thus, we cannot exclude the possibility that intermediate
duration detections (3-8 min) are also affected by neighbouring PMJ
activity. 
However, we decided to be conservative in manually sifting though the
detection statistics. 
Further, we note again our method's sensitivity to
weaker events which allow us to track events for longer duration as
compared to manual and by-eye selection methods.

\section{Conclusions and Summary \label{sec:conc}}

We studied PMJs using an automated simple machine learning detection 
scheme consisting of an initial Principle Component Analysis for the compression 
of data, and the subsequent application of the k-Nearest Neighbour algorithm and 
finally simple object tracking over the timeseries.
This scheme was applied to high spatial resolution observations of
well-sampled \ion{Ca}{ii}~8542 profiles.
We verified that the automated detections of PMJs in \ion{Ca}{ii}~8542
match well with PMJs in co-temporal \cahline core filtergrams, the
diagnostic for PMJs used in earlier studies. 
The \ion{Ca}{ii}~8542 PMJ line profile is characterized by enhanced
inner wings, often in the form of clear peaks, preferably with a
distinct asymmetry towards stronger blue wing enhancement. 
The line core is enhanced as compared to the quiet Sun reference
spectrum. 
We detect a total of 4253 PMJs at a detection rate of 21 events per 
timestep over a duration of 41 min corresponding to 453 PMJs tracked in time. 
Ellipse fitting to the PMJ detection areas yield average PMJ dimensions
of 640~km length and 210~km width. 
We measure an average lifetime of 90~s (discarding the longest
duration events, $>$8~min, that are clearly separate but overlapping
events in rapid succession or the result of misidentifications).
We detected PMJs in all parts of the penumbra, with many detections on 
both the limb-side as well as on the disk-center-side of the penumbra. 
However, there was still an apparent bias in that there were more 
detections on the limb- or upper- side of the sunspot, perhaps in part caused by 
foreshortening effects.
We note the existence of clear ``hot-spots'' with high occurrence
rates of PMJs.
\\
We finally remark that our results contribute to a solid observational
characterization of PMJs which is needed as constraints for theoretical and 
numerical modeling. Further research will necessarily have to focus on numerical 
studies to elucidate the precise physical nature of PMJs. Quantification of 
the heat-energy transfer by PMJs into the higher sunspot atmosphere is one such 
area of interest for future investigations.

\begin{acknowledgements}
The Swedish 1-m Solar Telescope is operated on the island of La Palma
by the Institute for Solar Physics of Stockholm University in the
Spanish Observatorio del Roque de los Muchachos of the Instituto de
Astrof{\'\i}sica de Canarias.
Our research has been partially funded by the Norwegian Research
Council and by the ERC under the European Union's Seventh Framework
Programme (FP7/2007-2013)\,/\,ERC grant agreement nr.~291058.
We made much use of NASA's Astrophysics Data System Bibliographic
Services.
\end{acknowledgements}

\bibliographystyle{aa}
\bibliography{biblio}

\end{document}